\documentclass[11pt]{article}
\usepackage{graphicx,graphicx,fancyhdr,mathrsfs}
\usepackage{amsfonts}
\usepackage{amsmath}
\usepackage{amssymb}
\usepackage{url}
\usepackage[colorlinks=true,citecolor=blue]{hyperref}
\usepackage{fancyhdr}
\usepackage{indentfirst}
\usepackage{enumerate}
\usepackage{amsthm}
\usepackage{color}
\usepackage{comment}
\usepackage{dsfont}
\usepackage{natbib}
\usepackage[misc]{ifsym}
\usepackage[toc,page]{appendix}
\usepackage{multirow}
\usepackage{float}
\usepackage{bm}
\usepackage{booktabs}
\usepackage{todonotes}
\usepackage{subcaption}
\usepackage{graphicx}
\usepackage{textcomp}

\def\d{\,\mathrm{d}}

\newcommand{\VaR}{\mathrm{VaR}}
\newcommand{\Var}{\mathrm{Var}}
\newcommand{\var}{\mathrm{var}}

\newcommand{\ES}{\mathrm{ES}}
\newcommand{\ex}{\mathrm{ex}}

\newcommand{\T}{\mathbb{T}}

\newcommand{\X}{\mathcal{X}}

\newcommand{\E}{\mathbb{E}}
\newcommand{\R}{\mathbb{R}}
\newcommand{\N}{\mathbb{N}}
\renewcommand{\L}{\mathcal{L}}
\newcommand{\M}{\mathcal{M}}

\newcommand{\p}{\mathbb{P}}

\newcommand{\id}{\mathds{1}}

\renewcommand{\ge}{\geqslant}
\renewcommand{\le}{\leqslant}

\theoremstyle{plain}
\newtheorem{theorem}{Theorem}

\newtheorem{proposition}{Proposition}
\theoremstyle{definition}

\newtheorem{example}{Example}

\newtheorem{assumption}{Assumption}

\newtheorem{remark}{Remark}

\DeclareMathOperator*{\argmin}{arg\,min}

\usepackage{tikz}

\renewcommand{\cite}{\citet}

\makeatletter
\DeclareRobustCommand{\bsquare}{%
	\mathop{\vphantom{\sum}\mathpalette\bigstar@\relax}\slimits@
}

\newcommand{\bigstar@}[2]{%
	\vcenter{%
		\sbox\z@{$#1\sum$}%
		\hbox{\resizebox{.9\dimexpr\ht\z@+\dp\z@}{!}{$\m@th\dsquare$}}%
	}%
}
\makeatother

\usepackage[onehalfspacing]{setspace}

\newcommand{\dsquare}{\mathop{  \square} \displaylimits}

\begin{document}
%\linenumbers

\title{Comparative e-backtesting}

\author{Zhanyi Jiao\thanks{School of Mathematical and Statistical Sciences, Arizona State University, U.S.A. \Letter~\url{zhanyij@asu.edu}} \and Qiuqi Wang\thanks{Maurice R.~Greenberg School of Risk Science, Georgia State University, U.S.A. \Letter~\url{qwang30@gsu.edu}} \and Yimiao Zhao\thanks{Royal Bank of Canada, Toronto, Ontario, Canada \Letter~\url{yimiao.zhao@rbccm.com}}}

 \date{\today}

\maketitle

\begin{abstract} 

Backtesting risk measures is a central task in financial regulation. While traditional backtests evaluate whether a forecasting model is statistically consistent with observed losses, regulatory practice often requires assessing the performance of an internal model relative to standard models.
In this paper, we develop a parametric anytime-valid framework for comparative backtests of general risk measures using e-values and e-processes.
We construct likelihood-based e-processes for comparative e-backtests with validity and power guarantees.
As practical guidelines, we propose the e-three-zone approach based on weak dominance, and introduce the overall and resetting test schemes for different practical purposes.
Simulation studies and empirical analyses are conducted to show practical manipulations in financial regulation.

\end{abstract}

\section{Introduction}
\label{sec:intro}

Forecasting risk measures like Value-at-Risk (VaR) and Expected Shortfall (ES) is central to modern financial risk management and regulation. Banks produce daily risk forecasts validated through backtesting, which provides statistical evidence of model performance and supports internal risk governance.
% \subsection{traditional backtest}
Traditional approaches, commonly referred to as \emph{traditional backtests}, evaluate whether a given forecasting model is statistically consistent with observed losses. Early studies such as \cite{K95} and \cite{C98} laid the foundation for VaR backtesting, which has become a core part of regulatory practice under Basel II and III \citep{B16}.

Backtesting ES is essentially more challenging because ES is not elicitable on its own \citep[as discussed in][]{G11}.
% As discussed in \cite{G11}, some risk measures, especially ES, are not elicitable on its own, meaning that there is no strictly consistent scoring function allowing for direct forecast evaluation. The non-elicitability of risk measure poses a fundamental challenge for model validation.
A number of studies explored possible remedies for ES, including joint elicitability of (VaR, ES) pairs \citep[][]{FZG16}, conditional calibration approaches \citep[][]{AS14}, expectile-based approaches for estimating VaR and ES \citep{T08}, and other related methods.
Recently, \cite{WWZ25} introduced a method for backtesting of VaR and ES based on e-values
\citep{S21,VW21,GHK24}.
% The feature of model-freeness (or non-parametric tests) for a backtesting approach is important for financial regulatory practice because the method is feasible without assuming specific distributions of financial losses, providing much practical flexibility given the complicated behavior of real losses. 
% \cite{WWZ25} achieve model-freeness using the tools of e-values and e-tests \citep{S21,VW21,GHK20}.
Compared with traditional p-value based tests commonly used in the existing literature, e-tests allow sequential any-time valid inference \citep[see e.g.,][]{RW25}.

% Building on this, a natural question arises: how can regulators adapt similar model-free backtesting methods to a broader class of risk measures beyond VaR and ES, such as expectiles \citep{NP87}? Inspired by \cite{CLZ22}, who constructed test supermartingales using identification functions for static tests, we aim to extend these ideas into a dynamic, sequential, e-value-based framework suited for regulatory practice.

% \subsection{Comparative backtest}

While traditional backtests assess whether a forecasting model is compatible with observed losses, they fail to evaluate the performance of an internal model relative to regulatory benchmarks, which is of practical interest. To overcome this, \emph{comparative backtests} were proposed \citep{FZG16, NZ17}, testing hypotheses
\begin{align*}
    & H_0^-: \text{The internal model is at least as accurate as the standard model,}\\
    & H_0^+: \text{The internal model is at most as accurate as the standard model.}
\end{align*}
Comparative backtests provide a more informative assessment of forecasting performance in regulatory settings, where internal models must be validated relative to prescribed benchmark models rather than evaluated in isolation. However, existing comparative backtesting procedures are subject to some limitations, leaving some important practical concerns unsolved:
\begin{enumerate}
    \item Financial losses and forecasts arrive sequentially, requiring procedures that remain valid under continuous monitoring and at data-dependent reporting dates. Fixed-horizon asymptotic tests do not provide this guarantee.
    % \item Score differences for ES and expectiles can be unbounded. Methods requiring bounded score differences or exponential-moment bounds do not directly apply without additional assumptions.
    \item An overall comparison can conceal changes in relative performance across market conditions over time. Repeated assessments over successive periods are therefore useful, but require error control across the reported periods.
\end{enumerate}

Financial regulatory comparative backtests have a different objective from a general procedure of comparing forecasts or a model selection problem \citep[see e.g.,][]{DM95}. Comparative backtests are designed for regulatory validation: the objective is to determine whether an internal model performs sufficiently well relative to a standard model under certain regulatory restrictions. This unique objective motivates the following features:
\begin{enumerate}
    \item Comparative backtests require standardized rules to determine whether an internal model passes or fails the test relative to the standard model instead of choosing the better model between both.
    \item Comparative backtests must be performed using a pre-speficied risk measure and scoring function satisfying the regulatory requirements. And unfortunately, most of the common scoring functions for ES and expectile do not have bounded differences, thus methods requiring bounded score differences do not apply.
    \item A regulator knows her own standard model, thus a comparative backtest is inherently not model-free. Depending on the quality of the standard model, the regulator can choose to take advantage of the additional information or not.
\end{enumerate}

In light of the limitations and features described above, in this paper, we develop a parametric sequential framework for comparative backtests of general risk measures based on the concept of e-tests.
The main contribution of this paper is three-fold.
\begin{enumerate}
    \item We develop a parametric sequential procedure for comparative e-backtests of general risk measures for financial regulatory practice (Section \ref{sec:com}). Specifically, we construct confidence regions using likelihood-based e-processes and report rejections based on the resulting range of the average score differences. The comparisons have finite-sample anytime validity (Theorem \ref{thm:adjusted_eprocess}) and power guarantee (Proposition \ref{prop:adjusted_power}). As a particular technical advantage, our construction of e-processes is valid without assuming boundedness of the score differences, which is a common assumption for established e-value-based sequential forecast comparison procedures \citep[e.g.,][]{CR24,AGSZ24}.\footnote{\cite{CR24,AGSZ24} also discuss unbounded cases as possible extensions: \cite{CR24} propose the asymptotic confidence sequence to handle the unbounded score differences, which loses finite-sample validity. \cite{AGSZ24} allow unbounded score differences under sub-Gaussian or sub-exponential conditions, which need not hold for heavy-tailed score differences.}
    \item We extend the three-zone approach of \cite{NZ17} to sequential comparisons based on weak dominance. The proposed e-three-zone approach reports mutually exclusive red and green conclusions for the same period. We propose overall and resetting test schemes with simultaneous directional error control and a uniform FDR guarantee (Section \ref{sec:schemes}), and discuss the choice of auxiliary densities and betting parameters (Section \ref{subsec:betting}).
    \item We demonstrate detailed comparative backtesting procedures for common risk measures, including VaR, ES and expectiles, for practical manipulations in financial regulation. In particular, we provide sufficient conditions for choosing auxiliary densities (Proposition \ref{prop:auxiliary_precision}) guaranteeing the test power. We demonstrate extensive numerical studies (Sections \ref{sec:simulation} and \ref{sec:real}) to illustrate the proposed procedures with simulated and real data.
\end{enumerate}

Additional technical details are presented in Appendix \ref{app:sec:tech}. Additional examples are provided in Appendix \ref{app:sec:ex}. Proofs of all results are put in Appendix \ref{app:sec:proofs}. Further numerical simulations and data analysis results are presented in Appendix \ref{app:sec:num}.

\subsection{Literature review}

Traditional backtests for various risk measures have been studied extensively in the literature. Following the seminal work of unconditional and independence tests conducted by \cite{K95} and \cite{C98}, the journey continues with backtesting VaR: \cite{EM04} studied backtesting conditional autoregressive VaR; \cite{EO10} considered the effect of estimation risk in backtesting problems; \cite{ZBWW14} backtest VaR with Monte-Carlo simulation; see \cite{BCP11} for a comprehensive overview of existing methods for backtesting VaR. Noticing the importance of ES in financial regulation, backtesting ES has also been a challenging but popular problem. \cite{MF00} and \cite{AS14} started by backtesting ES under independence; \cite{DE17} and \cite{SQPQ21} studied model-based backtests of ES using cumulative violations; \cite{BD22} introduced a linear regression-based method for ES backtests; \cite{HD23} designed a sequential p-test for monitoring ES; see \cite{WWZ25} for a comprehensive comparison of existing backtesting methods for ES, and their proposed model-free backtesting approach for ES based on e-values. Beyond VaR and ES, backtesting other elicitable risk measures is also of independent interest; see e.g., \cite{BNP19} for backtesting expectiles and \cite{CP17} for backtesting Lambda-Value-at-Risk.

Model selection and forecast comparison have long been important topics in economics and statistics. \cite{DM95} proposed the predictive ability test and forecast selection method to compare the predictive accuracy of competing forecasts. Their work was extended by \cite{CM01} and \cite{GW06}. \cite{HLN11} introduced the model confidence set method for model selection. \cite{CR24} developed a general sequential framework of comparing pairwise forecasts. \cite{AGSZ24} extended the model confidence set method into sequential testing. Under the framework of financial regulation, the term ``comparative backtest" was proposed by \cite{FZG16} and \cite{NZ17}. The comparative tests of \cite{NZ17} use fixed-horizon asymptotic inference, whereas our procedure provides finite-sample guarantees under sequential monitoring.

% Our comparative hypotheses differ from the calibration hypotheses studied by \cite{WWZ25}: a lower expected score relative to a standard forecast does not establish calibration. The resetting guarantee also differs from \cite{WWZ26}, where a report is false when all timewise null hypotheses in the reported interval hold. Here a false report gives the wrong direction of the interval's average conditional score difference. We retain the historical likelihood confidence sequence when restarting the score assessment and obtain error control over all reported intervals.

The major tools we are using are e-values and e-tests. The term ``e-value" is proposed by \cite{VW21}, whereas similar concepts can also be found in the early work of \cite{W45} and \cite{DR67}; the game-theoretic statistics and test martingales by \cite{SV19} and \cite{SSVV11}. E-values were applied to different key aspects of statistics: multiple testing problems \citep{VW21}; optimal criteria in sequential testing \citep{GHK24}; false discovery rate control \citep{WR22,RB24}; anytime-valid sequential testing problems of risk measures \citep{AKJ21,CLZ22,WWZ25}; sequential model selection procedures \citep{CR24,AGSZ24}. We refer to a review paper by \cite{RGVS23} for recent updates on e-values and the monograph by \cite{RW25} for a comprehensive summary of e-values.

\section{Preliminaries}
\label{sec:pre}

\subsection{Risk measures, elicitability, and identifiability}
\label{sec:elic}

We consider a probability space $(\Omega,\mathcal F,\mathbb P)$. Let $\X$ and $\M$ be some convex sets of random variables and distributions, respectively. Let $\X_+$  be the set of all non-negative random variables. For $n>0$, denote by $\L_n$ (resp.~$\M_n$) the set of random variables (resp.~distributions) with finite $n$-th moment. Let $\L_0$ and $\L_\infty$ (resp.~$\M_0$ and $\M_\infty$) be the sets of all random variables (resp.~distributions) and the sets of random variables (resp.~distributions) with bounded supports. For $x,y\in\R$, write $x\wedge y=\min\{x,y\}$, $x\vee y=\max\{x,y\}$, $x_+=x\vee 0$ and $x_-=-(-x)_+$. Let $\mathbb{T}$ be a finite or infinite set of indices starting from $0$, i.e., $\mathbb{T}=\{0,1,\dots,T\}$ for some $T\in\N\cup\{\infty\}$. Write $\mathbb{T}_+=\mathbb{T}\setminus\{0\}$.

For $d\in\N$, a \emph{risk measure} is a set-valued functional $\psi:\M\to 2^{\R^d}$. We write $\psi(X)=\psi(F)$ for a random variable $X\sim F\in\M$. Elicitability and identifiability are both important concepts for statistical inference. Following the definitions in \cite{FZ16}, a risk measure $\psi$ is called \emph{$\M$-elicitable} (or \emph{elicitable} for simplicity) if there exists a function $S:\R^{d+1}\to\R$ such that for all $F\in\M$,
$$\psi(F)=\argmin_{a\in\R^d}\int_\R S(x,a)\d F(x),$$
where $S$ is called a \emph{(strictly consistent) scoring function} for $\psi$ \citep{GR07,G11}. 
An integrable function $I:\R^{d+1}\to\R^d$ is said to be an \emph{$\M$-identification function} (or \emph{identification function} for simplicity) for $\psi$ if $\int_\R I(x,a)\d F(x)=\mathbf{0}$ for all $F\in\M$ and $a\in\psi(F)$.
Further, the risk measure $\psi$ is called \emph{$\M$-identifiable} (or \emph{identifiable} for simplicity) if there exists a \emph{strict} identification function $I:\R^{d+1}\to\R^d$ such that for all $F\in\M$ and $a\in\R^d$,
$$a\in\psi(F)\iff \int_\R I(x,a)\d F(x)=\mathbf{0}.$$
It is clear that given a true estimate $a$ of an elicitable (resp.~identifiable) risk measure $\psi$, it corresponds to the lowest expected score $S(\cdot,a)$ (resp.~zero expected identification $I(\cdot,a)$). This is the theoretical building block of a backtesting procedure.

Identifiable and elicitable risk measures are common in practice; see Appendix \ref{app:sec:ex} for examples and their corresponding identification and scoring functions.
In particular, fix a level $p\in(0,1)$. The \emph{lower-} and \emph{upper-Value-at-Risk} ($\VaR^-$ and $\VaR^+$) are defined as
$$\VaR^-_p(X)=\inf\{x\in\R:\p(X\le x)\ge p\}~~\mbox{and}~~\VaR^+_p(X)=\inf\{x\in\R:\p(X\le x)> p\},~~X\in \mathcal L_0,$$
respectively. The \emph{Value-at-Risk} (VaR) is then defined as a set-valued functional $\VaR_p(X)=[\VaR^-_p(X),\VaR^+_p(X)]$ for $X\in \mathcal L_0$. VaR is elicitable, and it is also identifiable for continuous distributions.
The \emph{Expected Shortfall} (ES) is defined as
$$\ES_p(X)=\frac{1}{1-p}\int^1_p \VaR^-_t(X)\,\d t,~~X\in \mathcal L_1.$$
ES is not itself identifiable or elicitable, whereas the pair $(\VaR,\ES)$ is both identifiable and elicitable for continuous distributions.
Another identifiable and elicitable common risk measure is the \emph{expectile} (ex), defined as
$$\ex_p(X)=\argmin_{a\in\R}\E\left[p(X-a)^2_++(1-p)(X-a)^2_-\right],~~X\in\L_2.$$

Throughout, we assume the risk measure $\psi:\M\to 2^{\R^d}$ is identifiable with a strict identification function $I:\R^{d+1}\to\R^d$, and is elicitable with a strictly consistent scoring function $S:\R^{d+1}\to\R$. All score differences and identification functions used below are assumed integrable under the considered probability measures and conditional loss distributions. The scoring and identification functions are considered on their stated domains, and all forecasts are assumed to lie in these domains.
% Following the insights of elicitability and Bayes risk, in the single-dimensional case, \cite{EMWW21} introduced Bayes pairs defined as follows. Let $I(\R)$ be the set of closed real intervals. A pair of risk measures $(\rho,\phi):\M\to \R\times I(\R)$ is a \emph{Bayes pair} if there exists a \emph{scoring} (or \emph{loss}) \emph{function} $S:\R^2\to\R$ such that for all $F\in\M$,
% $$\phi(F)=\argmin_{a\in\R}\int_\R S(x,a)\d F(x)~~\text{and}~~\rho(F)=\min_{a\in\R}\int_\R S(x,a)\d F(x).$$
% We first have the following lemma showing the link between Bayes pairs and identifiability, which will be useful for constructing our backtesting method.
% \begin{lemma}\label{lem:bayes}
%     Let $(\rho,\phi):\M\to\R\times I(\R)$ be a Bayes pair with a loss function $S:\R^2\to\R$. If $\phi$ has (strict) identification function $v:\R^2\to\R$, then $(\rho,\phi)$ has (strict) identification function $(x,r,z)\mapsto (v(x,z),h(r,z)(S(x,z)-r))^\top$ for all functions $h:\R^2\to\R$ which is not constantly $0$.
% \end{lemma}

\subsection{Traditional and comparative backtesting hypotheses}
\label{sec:null}

% We focus on dimensions $d=1$ or $2$ throughout to cover most common practical cases, whereas the cases of $d>2$ can be naturally extended from our framework.

Let $T\in\N\cup\{\infty\}$ be a fixed time horizon. Let $(L_t)_{t\in\T_+}$ be a sequence of random losses adapted to the filtration $(\mathcal F_t)_{t\in\T}$. We consider a \emph{predictable} process $(R_t)_{t\in\T_+}$, with $R_t:\Omega\to\R^d$ for $t\in\T_+$, forecasting the risk measure $(\psi(L_t|\mathcal F_{t-1}))_{t\in\T_+}$; that is, $(R_t)_{t\in\T_+}$ is adapted to $(\mathcal F_{t-1})_{t\in\T_+}$.
Using the terminology of \cite{NZ17}, the forecast sequence $(R_t)_{t\in\T_+}$ is said to be \emph{(conditionally) calibrated for $\psi$} if
\begin{align*}
% \label{eq:con_cali}
    \E\left[I(L_t,R_t)\mid\mathcal F_{t-1}\right]=0~~\mbox{a.s., for all}~t\in\T_+.
\end{align*}
% It is further called \emph{conditionally super-(resp.~sub-)calibrated for $\psi$} by changing ``$=$" to ``$\ge$" (resp.~``$\le$") in \eqref{eq:con_cali}.
A \emph{traditional backtest} therefore studies the following null hypothesis:
\begin{align}\label{eq:trad_h0}
    H(\psi):~(R_t)_{t\in\T_+}~\mbox{is calibrated for}~\psi.
\end{align}
By definition, traditional backtests detect whether the predictions reported by the financial institution deviate too much from the correct model. However, correct models are usually unknown in practice, and the financial regulators only have some standard models. In order to compare several internal models and the standard models they have, regulators resort to comparative backtests.

% Both unconditional and conditional notions of predictive ability were adopted in existing literature. \cite{DM95} and \cite{NZ17} used unconditional score comparisons for predictive ability tests and comparative backtests, while \cite{GW06} studied conditional predictive ability. Sequential comparisons were considered by \cite{HZ22,CR24}.

Let $(R_t)_{t\in\T_+}$ denote the internal model provided by the financial institution and $(R^*_t)_{t\in\T_+}$ denote the standard model of the regulator, both predictable and $R_t,R^*_t:\Omega\to\R^d$ for $t\in\T_+$. We can compare the internal and standard models in the following sense: the forecast $(R_t)_{t\in\T_+}$ is said to \emph{weakly (conditionally) dominate} $(R^*_t)_{t\in\T_+}$ if
\begin{equation}\label{eq:w_dom}
    \frac{1}{t}\sum^t_{s=1}\E\left[S(L_s,R_s)-S(L_s,R^*_s)\mid\mathcal F_{s-1}\right]\le 0~~\mbox{a.s., for all}~t\in\T_+.
\end{equation}
Our notion of weak $S$-dominance is in the same sense as the weak hypothesis of \cite{CR24} and the uniform weak superiority of \cite{AGSZ24}. It is weaker than conditional $S$-dominance defined by \cite{NZ17}. The $S$-dominance on average of \cite{NZ17} instead requires $\E[S(L_t,R_t)-S(L_t,R_t^*)]\le0$ for every $t$; neither this condition nor \eqref{eq:w_dom} implies the other in general.
In light of \eqref{eq:w_dom}, we study a \emph{comparative backtest} for the following hypotheses:
\begin{align}
    &H^-(\psi):~(R_t)_{t\in\T_+}~\mbox{weakly dominates}~(R^*_t)_{t\in\T_+},\label{eq:com_h0-}\\
    &H^+(\psi):~(R_t)_{t\in\T_+}~\mbox{is weakly dominated by}~(R^*_t)_{t\in\T_+}.\label{eq:com_h0+}
\end{align}
It is necessary to consider two hypotheses for a comparative backtest in practice because if a single hypothesis $H^-(\psi)$ is not rejected throughout, we are not able to say the internal model passes the comparative backtest without further justification by rejecting $H^+(\psi)$. We test weak dominance instead of the stronger conditional $S$-dominance for two reasons:
\begin{enumerate}
    \item Our tests of weak dominance yield mutually exclusive rejections of the two hypotheses for the same assessment period at each time $t$ (see Section \ref{sec:traffic} below). This makes the result easier to interpret for practitioners.
    \item The main goal for our comparative backtests is to sequentially assess the average performance of the internal model versus the standard model up to each time $t$. On the other hand, a test on conditional $S$-dominance raises alarms for the internal model being conditionally worse than the standard model at some date in the reported period, which does not fit our purpose.
\end{enumerate}

\subsection{E-values and e-processes}
\label{sec:e}

We develop our backtesting approach using the concept of e-values and e-processes. Let $H$ be a set of probability measures on $(\Omega,\mathcal F)$, representing the (composite) hypothesis we are testing. The hypothesis $H$ is simple if it is a singleton. Following the terminology of \cite{VW21}, we say a non-negative random variable $E$ is an \emph{e-variable} for $H$ if $\E^Q[E]\le 1$ for all $Q\in H$. A realized e-variable is called an \emph{e-value}. An \emph{e-test} for $H$ rejects $H$ when an e-value exceeds a prespecified threshold. In a dynamic setup, we say a non-negative stochastic process $(M_t)_{t\in K}$, $K\subseteq \N$, adapted to a filtration $(\mathcal F_t)_{t\in K}$, is an \emph{e-process} for $H$ if $\E^Q[M_\tau]\le 1$ for any stopping time $\tau$ taking values in $K$ and all $Q\in H$. A process can be constructed by a supermartingale. Namely, for $\mathcal F_t$-measurable e-values $E_t$ with $\E^Q[E_t|\mathcal F_{t-1}]\le 1$ for all $Q\in H$ and a predictable process $(\lambda_t)_{t\in\T_+}$ taking values in $[0,1]$, the process $(M_t)_{t\in\T}$ with $M_0=1$ and
\begin{equation}\label{eq:mtg0}
    M_t=M_{t-1}(1-\lambda_t+\lambda_tE_t)=\prod^t_{s=1}(1-\lambda_s+\lambda_sE_s),~~t\in\T_+,
\end{equation}
is a supermartingale under all $Q\in H$, and is an e-process for $H$. It is clear by definition that the e-process in \eqref{eq:mtg0} satisfies optional stopping and guarantees the following uniform type-I error control by Ville's inequality:
\begin{equation*}
    Q\left(\sup_{t\in\T_+} M_t\ge 1/\alpha\right)\le \alpha~~\mbox{for all}~\alpha\in(0,1)~\mbox{and}~Q\in H.
\end{equation*}

One can easily construct e-values and e-processes with scoring and identification functions. We take an example similar to the construction of \cite{CLZ22}. Suppose that the losses $(L_t)_{t\in\T_+}$ take values in a measurable space $\mathcal K$. Fix $t\in\T_+$. For predictable $h_t\in\R^d$ satisfying the uniform boundedness condition $\inf_{x\in\mathcal K,a\in\psi(\M)}h^{\top}_tI(x,a)\ge -1$, we can take
\begin{equation}\label{eq:naive_e1}
    E_t=1+h^{\top}_tI(L_t,R_t)
\end{equation}
as an $\mathcal F_t$-measurable e-variable for $H(\psi)$ in \eqref{eq:trad_h0}. This is essentially a backtest e-statistic studied by \cite{WWZ25} for traditional backtests. See Appendix \ref{app:sec:iden} for identification functions of common risk measures and their corresponding uniform bounds.
Similarly, for conditional $S$-dominance, let predictable $\lambda^-_t,\lambda^+_t\ge0$ be such that
\begin{equation*}
\begin{aligned}
    \inf_{x\in\mathcal K}\lambda^-_t(S(x,R_t)-S(x,R_t^*))\ge-1~~\mbox{and}~~\inf_{x\in\mathcal K}\lambda^+_t(S(x,R_t^*)-S(x,R_t))\ge-1.
\end{aligned}
\end{equation*}
We can take
\begin{equation}\label{eq:naive_e2}
    \begin{aligned}
        E_t^-=1+\lambda^-_t\left(S(L_t,R_t)-S(L_t,R_t^*)\right)~~\mbox{and}~~E_t^+=1+\lambda^+_t\left(S(L_t,R_t^*)-S(L_t,R_t)\right)
    \end{aligned}
\end{equation}
as $\mathcal F_t$-measurable e-variables for conditional $S$-dominance in the respective directions. Under their respective conditional hypotheses, plugging \eqref{eq:naive_e1} or \eqref{eq:naive_e2} into \eqref{eq:mtg0} gives the corresponding e-processes. Comparative e-backtests for the weak-dominance hypotheses \eqref{eq:com_h0-} and \eqref{eq:com_h0+} are constructed in Section \ref{sec:com_evalue} below.

\section{Comparative e-backtests}
\label{sec:com}

In this section, we demonstrate our main methodology incorporating the practical features of comparative backtests. We call our procedure a \emph{comparative e-backtest}.

\subsection{Constructing e-processes using likelihood ratios}
\label{sec:com_evalue}

Write the average expected score difference as
\begin{equation}\label{eq:assessment_process}
    \Delta_{s,t}=\frac1{t-s}\sum_{u=s+1}^t\E[S(L_u,R_u)-S(L_u,R_u^*)\mid\mathcal F_{u-1}],
    \qquad \Delta_t=\Delta_{0,t},\qquad 0\le s<t.
\end{equation}
The hypotheses in \eqref{eq:com_h0-}--\eqref{eq:com_h0+} can then be written as
\begin{align}
    H^-(\psi)&:\quad \Delta_t\le0\quad \p\text{-a.s.~for all }t\in\T_+,\label{eq:model_com_h0-}\\
    H^+(\psi)&:\quad \Delta_t\ge0\quad \p\text{-a.s.~for all }t\in\T_+.\label{eq:model_com_h0+}
\end{align}
A practical challenge for constructing score-based e-processes is that the score difference $S(L_t,R_t)-S(L_t,R_t^*)$ may be unbounded. Established work mostly focuses on pre-assumed bounded score differences \citep[e.g.,][]{CR24,AGSZ24}. However, such bounds no longer hold for most of the score differences used for common risk measures in comparative backtests; see Appendices \ref{app:sec:score1} and \ref{app:sec:score0} for details. Even for bounded losses, a bound based on a large prescribed loss range may lead to conservative tests. 
Our proposed procedure remains valid for general possibly unbounded score differences.
We solve the boundedness issue from a different perspective by constructing e-processes using likelihood ratios in a general parametric setup, where validity still holds for unbounded score differences.

Let $Q_{t,\theta}$ be a predictable parametric family of conditional loss distributions with strictly positive densities $q_{t,\theta}$, indexed by a common unknown finite-dimensional parameter $\theta\in\Theta$. We assume that
\begin{equation}\label{eq:auxiliary_correction}
    \p(L_t\in\mathrm dx\mid\mathcal F_{t-1})
       =Q_{t,\theta_0}(\mathrm dx)
       =q_{t,\theta_0}(x)\,\mathrm dx,
       \qquad t\in\T_+,
\end{equation}
for some $\theta_0\in\Theta$. For initialized time-series models, $\theta_0$ may include an unknown $\mathcal F_0$-measurable initial state.
Choose a predictable auxiliary probability density $q_t$, which may be fitted using past observations. In a similar sense to the sequential likelihood construction of \cite{WRB20}, for predictable $\lambda_t\in[0,1]$ define $E_0(\theta)=1$ and
\begin{equation}\label{eq:likelihood_parameter_eprocess}
    E_t(\theta)=\prod_{u=1}^t
       \left(1-\lambda_u+\lambda_u
          \frac{q_u(L_u)}{q_{u,\theta}(L_u)}\right),
       \qquad t\in\T_+.
\end{equation}
For each fixed $\theta\in\Theta$, $(E_t(\theta))_{t\in\T}$ is a non-negative martingale, and hence a valid e-process, for the hypothesis \[     \p(L_t\in\mathrm dx\mid\mathcal F_{t-1})        =Q_{t,\theta}(\mathrm dx),\qquad t\in\T_+. \]
We call $\lambda_t\in[0,1]$ in \eqref{eq:likelihood_parameter_eprocess} a \emph{betting parameter} for the e-processes. Each $E_t(\theta)$ gives an e-value at a reporting date under the corresponding conditional-model hypothesis.
In practice, one can choose $q_t$ from a benchmark auxiliary distribution. The auxiliary density $q_t$ is allowed to be mis-specified, which does not affect validity of the e-process.

For a pre-specified level $\alpha\in(0,1)$, set the confidence region of parameters for which the e-process does not exceed threshold $1/\alpha$:
\begin{equation}\label{eq:likelihood_parameter_region}
    \mathcal C_{t}
       =\{\theta\in\Theta:E_{t}(\theta)<1/\alpha\},~~t\in\T.
\end{equation}
For parameter $\theta\in\Theta$, write the corresponding average expected score difference under probability measures $(Q_{u,\theta})_{u=s+1,\dots,t}$ as
\[
    \Delta_{s,t}(\theta)=\frac1{t-s}\sum_{u=s+1}^t
       \int_\R\{S(x,R_u)-S(x,R_u^*)\}\d Q_{u,\theta}(x).
\]
At each reporting date $t\in\T_+$, calculate the upper and lower bounds of the score differences over the confidence region $\mathcal C_{t-1}$:
\begin{equation}\label{eq:auxiliary_projection}
    \underline\Delta_{s,t}=\inf_{\theta\in\mathcal C_{t-1}}\Delta_{s,t}(\theta),
    \qquad
    \overline\Delta_{s,t}=\sup_{\theta\in\mathcal C_{t-1}}\Delta_{s,t}(\theta),
\end{equation}
where we assume $\underline\Delta_{s,t}$ and $\overline\Delta_{s,t}$ are $\mathcal F_{t-1}$-measurable. If $\mathcal C_{t-1}=\varnothing$, set $\underline\Delta_{s,t}=-\infty$ and $\overline\Delta_{s,t}=\infty$.
For numerical implementation, let $\widehat{\mathcal C}_{t-1}$ denote the approximation of $\mathcal C_{t-1}$ used to evaluate \eqref{eq:auxiliary_projection}; see Appendix \ref{app:sec:confidence_computation} for its construction.

The following result shows the validity of our comparative e-backtests through a type-I error control.

\begin{theorem}\label{thm:adjusted_eprocess}
    Suppose \eqref{eq:auxiliary_correction} holds. We have
    \begin{equation}\label{eq:auxiliary_cs}
        \p(\theta_0\in\mathcal C_{t}\text{ for all }t\in\T)\ge1-\alpha,
    \end{equation}
    and
    \begin{equation}\label{eq:directional_validity}
    \p\left(\exists\,0\le s<t:
       \{\underline\Delta_{s,t}>0,\ \Delta_{s,t}\le0\}
       \cup
       \{\overline\Delta_{s,t}<0,\ \Delta_{s,t}\ge0\}\right)
       \le\alpha.
    \end{equation}
\end{theorem}
The bound is simultaneous over both directions and all assessment periods, including periods starting or ending at stopping times. In other words, the testing procedure is anytime valid. Moreover, the next result shows the power guarantee of the proposed procedure.
% For the overall test ($s=0$), it gives level-$\alpha$ anytime tests of \eqref{eq:model_com_h0-} and \eqref{eq:model_com_h0+} in the respective directions.

\begin{proposition}\label{prop:adjusted_power}
    Let $T=\infty$. Suppose that
    \begin{equation}\label{eq:prop1_conv}
        \underline\Delta_{0,t}-\Delta_t\xrightarrow{\mathrm{p}} 0~~\text{and}~~\overline\Delta_{0,t}-\Delta_t\xrightarrow{\mathrm{p}} 0~~\mbox{as}~t\to\infty.
    \end{equation}
    For a constant $\varepsilon>0$,
    \[
    \begin{aligned}
        \liminf_{t\to\infty}\Delta_t\ge\varepsilon~~\text{a.s.}
        &\implies
        \p(\underline\Delta_{0,t}>0)\to 1~~\mbox{as}~t\to\infty,\\
        \limsup_{t\to\infty}\Delta_t\le-\varepsilon~~\text{a.s.}
        &\implies
        \p(\overline\Delta_{0,t}<0)\to 1~~\mbox{as}~t\to\infty.
    \end{aligned}
    \]
    If both convergences in \eqref{eq:prop1_conv} hold a.s., then
    \[
    \begin{aligned}
        \liminf_{t\to\infty}\Delta_t\ge\varepsilon~~\text{a.s.}
        &\implies
        \p(\exists\,n\in\N:\ \underline\Delta_{0,t}>0,\ \forall t\ge n)=1,\\
        \limsup_{t\to\infty}\Delta_t\le-\varepsilon~~\text{a.s.}
        &\implies
        \p(\exists\,n\in\N:\ \overline\Delta_{0,t}<0,\ \forall t\ge n)=1.
    \end{aligned}
    \]
\end{proposition}
The power guarantee in Proposition \ref{prop:adjusted_power} requires the convergences of the bounds of the average expected score differences as in \eqref{eq:prop1_conv}. This is affected by the choice of the auxiliary density $q_t$ in the e-process \eqref{eq:likelihood_parameter_eprocess}; see Section \ref{subsec:betting} below for proper choices of the auxiliary densities with power guarantee.

\subsection{The e-three-zone approach for comparative e-backtests}
\label{sec:traffic}

In light of the e-processes, validity, and power constructed in Theorem \ref{thm:adjusted_eprocess} and Proposition \ref{prop:adjusted_power}, we modify the three-zone approach following \cite{FZG16,NZ17} for practical use, which we call an \emph{e-three-zone approach}.

For conditional $S$-dominance, both hypotheses may be false when the internal model is conditionally better than the standard model at some dates and conditionally worse at others. We instead compare average performance over the same assessment period, as in \eqref{eq:assessment_process}.

For an assessment $(s,t]$, the comparative e-backtest reports:
\begin{enumerate}[(i)]
    \item \emph{Red} if $\underline\Delta_{s,t}>0$. There is evidence that the internal model has a higher average conditional score than the standard model and does not pass the comparative backtest.
    \item \emph{Green} if $\overline\Delta_{s,t}<0$. There is evidence that the internal model has a lower average conditional score than the standard model and passes the comparative backtest.
    \item \emph{Yellow} otherwise. No clear conclusion is obtained.
\end{enumerate}
The red and green regions are mutually exclusive because $\underline\Delta_{s,t}\le\overline\Delta_{s,t}$. Under \eqref{eq:auxiliary_correction}, \eqref{eq:directional_validity} controls the probability of ever reporting red when $\Delta_{s,t}\le0$ or green when $\Delta_{s,t}\ge0$ at level $\alpha$. Yellow does not establish equal performance or that either model is correct.

Exchanging the models negates the raw score difference and interchanges red and green when the same likelihood confidence region is used. Given the important feature of anytime validity of our testing procedure, the color can change sequentially as new data arrive. Opposite rejections at different dates concern different evaluation periods.

\subsection{Overall and resetting test schemes}
\label{sec:schemes}

We propose two test schemes for different regulatory purposes. An overall test evaluates the average performance of the internal model over a prespecified period. On the other hand, a resetting test reports successive time intervals with evidence of inferior or superior performance. This is of interest if the regulator is more concerned about the dynamics of the performance of the internal model over time. Specifically, suppose that the internal model outperforms the standard model in regular periods but poorly performs during a crisis. Our framework allows the conclusion to change over time. This can be achieved by restarting the score assessment at certain points, while each report concerns average performance over its own interval.
\begin{enumerate}[(i)]
    \item \emph{Overall test.} Calculate \eqref{eq:auxiliary_projection} with $s=0$ and report the color in Section~\ref{sec:traffic} at any reporting date, which may be a stopping time. For a prespecified final reporting date $\tau$, use the terminal interval; an earlier colored report does not require stopping the test. For an assessment starting at $s>0$, replace $\Delta_t$ in \eqref{eq:model_com_h0-}--\eqref{eq:model_com_h0+} by $\Delta_{s,t}$ for all $t>s$, using the same confidence sequence.
    \item \emph{Resetting test.} Let $\tau_0=0$ and define
    \begin{equation}\label{eq:active_reset}
        \tau_k=\inf\{t>\tau_{k-1}:
        \underline\Delta_{\tau_{k-1},t}>0
        \text{ or }\overline\Delta_{\tau_{k-1},t}<0\},
    \end{equation}
    with $\inf\emptyset=\infty$. At a finite $\tau_k$, report red or green for $(\tau_{k-1},\tau_k]$ and begin the next assessment at the following observation. If $\tau_{k-1}=\infty$, set $\tau_k=\infty$.
\end{enumerate}
Each resetting report compares only observations since the previous restart. Forecast estimation and the likelihood confidence sequence continue to use the available history.
The two colors remain mutually exclusive within each assessment. A reset changes the period being evaluated, so it may generate rejection intervals with varying colors.

Both the overall and resetting test schemes are theoretically justified: Under \eqref{eq:auxiliary_correction}, Theorem~\ref{thm:adjusted_eprocess} gives the following validity guarantees for the overall test:
\[
\begin{aligned}
    H^-(\psi)\implies
       \p(\exists\,t\in\T_+:\underline\Delta_{0,t}>0)\le\alpha,~~~H^+(\psi)\implies
       \p(\exists\,t\in\T_+:\overline\Delta_{0,t}<0)\le\alpha.
\end{aligned}
\]
Because the resetting test generates multiple reports of intervals $(\tau_{k-1},\tau_k]$ across successive assessments $k$, it involves a multiple testing problem, where false discovery rate control \citep{BH95} is an important task. A completed report for $(\tau_{k-1},\tau_k]$ is \emph{false} if it is red with $\Delta_{\tau_{k-1},\tau_k}\le0$ or green with $\Delta_{\tau_{k-1},\tau_k}\ge0$. The numbers of completed reports and false reports by time $t$ are
\[
    N_t=\sum_{k\in\N}\id_{\{\tau_k\le t\}},\qquad
    V_t=\sum_{k\in\N}\id_{\{k\text{-th report is false}\}}
                       \id_{\{\tau_k\le t\}}.
\]
Under \eqref{eq:auxiliary_correction}, Theorem~\ref{thm:adjusted_eprocess} also bounds the probability of ever issuing a false resetting report:
\[
    \p(\exists\,t\in\T:V_t\ge1)\le\alpha.
\]
The \emph{false discovery rate} (FDR) and \emph{uniform FDR} are defined respectively as
\[
    \mathrm{FDR}_t=\E\left[\frac{V_t}{N_t\vee1}\right],
    \qquad
    \mathrm{FDR}_\infty=\E\left[\sup_{t\in\T}
                         \frac{V_t}{N_t\vee1}\right].
\]
The uniform FDR guarantee in \cite[Theorem 2]{WWZ26} is established for test supermartingales, with a report defined as false when all timewise null hypotheses in the reported interval hold. In our framework, false reports concern the direction of the average conditional score difference, and an FDR guarantee is not implied by \cite{WWZ26}. We present the following uniform FDR bound as an independent result with a separate proof based on Theorem \ref{thm:adjusted_eprocess}.

\begin{proposition}\label{prop:reset_fdr}
    Under the conditions of Theorem~\ref{thm:adjusted_eprocess}, the resetting scheme \eqref{eq:active_reset} satisfies
    $\mathrm{FDR}_\tau\le\mathrm{FDR}_\infty\le\alpha$
    for every stopping time $\tau$ taking values in $\T_+$.
\end{proposition}
Proposition \ref{prop:reset_fdr} provides guidance for choosing rejection thresholds for resetting tests, where we guarantee an FDR below $\alpha$ by a fixed rejection threshold of $1/\alpha$. This FDR bound is lower than the general bound $\alpha(1-\log(\alpha))$ obtained by \cite{WWZ26}.

\subsection{Choosing the betting parameter and auxiliary density}
\label{subsec:betting}

In this section, we specify how to choose the betting parameter $\lambda_t$ and the auxiliary density $q_t$ in practice.

We use $\lambda_t=1/2$ in the numerical studies to bound each likelihood factor below by $1/2$, thus a single observation cannot reduce capital by more than half. Predictable choices $\lambda_t\in[0,1]$ also retain validity. For adaptive choices based on empirical log growth, see e.g., \cite{WSR24,WWZ25}.

We recall that the choice of density $q_t$ does not affect validity of the e-test but does affect power. The auxiliary density $q_t$ predicts the full conditional distribution of $L_t$ and enters the likelihood ratio in \eqref{eq:likelihood_parameter_eprocess}. Its predictive accuracy is measured by the log loss $-\log q_t(L_t)$. This is more demanding than producing correctly calibrated risk forecasts, which are compared using the scoring function $S$.
For a practical choice of auxiliary density, we use expanding conditional likelihood. The following result shows a sufficient condition of the power guarantee for a proper choice of the auxiliary density.

\begin{proposition}\label{prop:auxiliary_precision}
    Let $T=\infty$. Suppose \eqref{eq:auxiliary_correction} and Assumption \ref{app:assum:stationary} in Appendix \ref{app:sec:assum} hold. Fix $\lambda_t=\lambda\in(0,1)$. Let the auxiliary density be $q_t=q_{t,\widehat\theta_{t-1}}$, where $\widehat\theta_{t-1}\in\Theta$ is $\mathcal F_{t-1}$-measurable for every $t\in\T_+$ and satisfies
        \begin{equation}\label{eq:auxiliary_fit}
            \sup_{\theta\in\Theta}\sum_{u=1}^t
               \log\frac{q_{u,\theta}(L_u)}{q_{u,\widehat\theta_t}(L_u)}=o(t)~~\mbox{a.s.}
        \end{equation}
    If
        \begin{equation}\label{eq:auxiliary_envelope}
        \begin{aligned}
         &\E\!\left[\sup_{\theta\in\Theta}|\log q_{1,\theta}(L_1)|\right]<\infty,\\
         \mbox{and}~~&\E\!\left[\sup_{\theta\in\Theta}\int_\R
            |S(x,R_1)-S(x,R_1^*)-S(0,R_1)+S(0,R_1^*)|
                \d Q_{1,\theta}(x)\right]<\infty,
        \end{aligned}
        \end{equation}
    then both differences in \eqref{eq:prop1_conv} converge to zero simultaneously with probability at least $1-\alpha$, and with probability one if $\mathcal C_{t-1}\ne\varnothing$ for all sufficiently large $t$, a.s.
\end{proposition}

Condition \eqref{eq:auxiliary_fit} holds for the conditional maximum likelihood estimator, or an approximate maximizer whose log-likelihood gap is $o(t)$ a.s.

For location--scale families, write $Q_{u,\theta}$ as the law of $\mu_u(\theta)+\sigma_u(\theta)Z_\theta$. Suppose the locations $\mu_u(\theta)$ and scales $\sigma_u(\theta)$ are continuous in $\theta$ a.s.~and the law of innovation $Z_\theta$ is weakly continuous in $\theta$. Explicit sufficient conditions for the score continuity in Assumption \ref{app:assum:stationary} and the second bound in \eqref{eq:auxiliary_envelope} are (see proof of Proposition \ref{prop:auxiliary_precision})
\begin{equation}\label{eq:raw_location_scale_precision}
\begin{aligned}
 \E\!\left[\operatorname{Lip}\{S(\cdot,R_1)-S(\cdot,R_1^*)\}^{2}\right]<\infty,~~\E\!\left[\sup_{\theta\in\Theta}
            \{|\mu_1(\theta)|^2+\sigma_1(\theta)^2\}\right]<\infty,
 ~~\sup_{\theta\in\Theta}\E|Z_\theta|^2<\infty.
\end{aligned}
\end{equation}
Here $\operatorname{Lip}$ denotes a Lipschitz constant. For the scores used in our numerical studies, the first condition in \eqref{eq:raw_location_scale_precision} can be checked as follows:
\begin{center}
\begin{tabular}{ll}
\hline
Score & Sufficient forecast condition\\
\hline
$\VaR_p$ & Automatic\\
$\ex_p$ & $\E|R_1-R_1^*|^2<\infty$\\
$(\ES_p,\VaR_p)$ & $R^{(1)}_1,R^{(1)*}_1>0$, $\E(1/R^{(1)}_1\vee1/R^{(1)*}_1)<\infty$\\
\hline
\end{tabular}
\end{center}
Here $R^{(1)}_u,R^{(1)*}_u$ are the ES forecasts in \eqref{eq:scoreVaRES}.
For centered variance-one innovations, the third condition in
\eqref{eq:raw_location_scale_precision} holds automatically.
Examples include standard normal and variance-standardized Student
innovations with $\nu>2$. See Appendix~\ref{app:sec:innovation_densities} for the Student and skewed-t densities. The second condition in \eqref{eq:raw_location_scale_precision}
is directly verifiable for specific conditional models. For the
stationary AR--GARCH model underlying our numerical studies, prespecify $0<\varepsilon<1$ and $B>2+\varepsilon$, and use a closed structural parameter domain satisfying
\begin{equation}\label{eq:argarch_power_bounds}
\begin{aligned}
 |\phi_0|\le B,~~ |\phi_1|\le1-\varepsilon,
       ~~ \varepsilon\le\alpha_0\le B,~\alpha_1,\beta_1\ge0,~~\alpha_1+\beta_1\le1-\varepsilon,~~2+\varepsilon\le\nu\le B,~~ \varepsilon\le\gamma\le B.
\end{aligned}
\end{equation}
The second condition in \eqref{eq:raw_location_scale_precision} follows from \eqref{eq:argarch_power_bounds} and a finite innovation variance.
Appendix~\ref{app:sec:argarch} extends the result to an unknown
finite initial variance without a prespecified upper bound.

\begin{remark}
In practice, the regulator usually knows her standard model. However, the auxiliary distribution need not be calibrated to the standard forecast or share its fitted parameters. The known standard density can itself be used as $q_t$, depending on the regulator's confidence in the quality of her standard forecasts.
There is another potential route to improve test power given the information of a properly calibrated standard model: if the regulator can justify prespecified predictable tolerance levels $\delta_u^-,\delta_u^+\in[0,\infty]^d$ satisfying
\[
    -\delta_u^-\le\E[I(L_u,R_u^*)\mid\mathcal F_{u-1}]
       \le\delta_u^+\quad\mbox{a.s.~for all }u\in\T_+,
\]
componentwise, the likelihood region may be narrowed by
\begin{equation*}
    \mathcal C_{t-1}\cap
    \left\{\theta\in\Theta:
       -\delta_u^-\le\int_\R I(x,R_u^*)\d Q_{u,\theta}(x)
          \le\delta_u^+\text{ for every }u\le t\right\},
\end{equation*}
when it is non-empty.
Let $\delta=(\delta^-_u,\delta^+_u)_{u\in\T_+}$. It is straightforward that the resulting upper bound $\overline\Delta_{s,t}(\delta)\le \overline\Delta_{s,t}$ and the lower bound $\underline\Delta_{s,t}(\delta)\ge \underline\Delta_{s,t}$. Thus improving the test power with guaranteed validity if the calibration tolerance levels $\delta^-_u$ and $\delta^+_u$ are truly specified. Infinite tolerances leave the likelihood region unchanged.
\end{remark}

\section{Simulation studies}
\label{sec:simulation}

In this section, we present simulation studies for VaR, ES, and expectile. We use the $1$-homogeneous VaR, squared expectile, and $1/2$-homogeneous VaR--ES scoring functions in Example \ref{ex:2}. For the likelihood tests, we use \eqref{eq:likelihood_parameter_eprocess} with $\lambda_t=1/2$ and threshold $20$, giving $\alpha=0.05$. The confidence regions in \eqref{eq:likelihood_parameter_region} are numerically approximated; see Appendix \ref{app:sec:confidence_computation} for details. 
For overall tests, we present the results as tables and heatmaps for the e-three-zone approach at the final assessment date. For resetting tests, we show the results of the e-three-zone approach with resetting colored rejection blocks. Exchanging the internal and standard models interchanges red and green.

\subsection{Type-I error and test power for comparative e-backtests}
\label{sec:typeI}

To examine the type-I error and power, we design a simple data setup with clear dominance relations. We simulate $l=10$ observations for training and $n=1000$ observations for testing from $L\sim\mathrm{N}(0,1)$ and compare six reporting strategies for $\VaR_{0.95}(L)$: (i) perfect information with $R_t=\VaR_{0.95}(L)\approx1.64$, (ii) truth reporting subject to noise, $R_t=\VaR_{0.95}(L)+\varepsilon_t$, where $\varepsilon_t$ are iid uniform on $\{\pm i/20:i=0,\dots,5\}$, and (iii) under- or over-reporting VaR plus the same noise, $R_t=(1+\theta)\VaR_{0.95}(L)+\varepsilon_t$, where $\theta=\pm 20\%$ or $\pm 50\%$. We use the same noise sequence in all 1000 repetitions and record the average performance.

For constructing the test, we treat the mean and standard deviation of the normal loss distribution as unknown. At each date \(t>l\), we estimate them by maximum likelihood using \(L_1,\ldots,L_{t-1}\), and use the fitted normal density as \(q_t\). We do an overall test, where rejection decisions are made at the terminal time $T=l+n$ according to $\underline{\Delta}_{l,T}>0$ for red, $\overline{\Delta}_{l,T}<0$ for green, and yellow otherwise using the e-three-zone approach.

\begin{table}[htbp]
\centering
\footnotesize
\renewcommand{\arraystretch}{1.3}
\resizebox{0.90\textwidth}{!}{\begin{tabular}{@{}lrrrrrr@{}}
\toprule
Standard / Internal & -50\% & -20\% & truth & +20\% & +50\% & opt \\
\midrule
-50\% & -- & 0.0/100.0 & 0.0/100.0 & 0.0/100.0 & 0.0/76.0 & 0.0/100.0 \\
-20\% & 100.0/0.0 & -- & 0.0/65.2 & 0.0/0.3 & 24.9/0.0 & 0.0/85.5 \\
truth & 100.0/0.0 & 65.2/0.0 & -- & 21.2/0.0 & 100.0/0.0 & 0.0/100.0 \\
+20\% & 100.0/0.0 & 0.3/0.0 & 0.0/21.2 & -- & 100.0/0.0 & 0.0/51.9 \\
+50\% & 76.0/0.0 & 0.0/24.9 & 0.0/100.0 & 0.0/100.0 & -- & 0.0/100.0 \\
opt & 100.0/0.0 & 85.5/0.0 & 100.0/0.0 & 51.9/0.0 & 100.0/0.0 & -- \\
\bottomrule
\end{tabular}
}
\caption{Red/green rejection rates (\%) for VaR forecasts of the iid normal simulation over 1000 runs. ``truth'' represents the truth reporting subject to noise and ``opt'' represents the forecasts with perfect information. Columns represent internal models and rows represent standard models.}
\label{tab:typeI}
\end{table}

Table \ref{tab:typeI} presents the terminal rejection rates. Under-reporting by $20\%$ and $50\%$ is rejected against the perfect-information standard model in $85.5\%$ and $100\%$ of the runs. The corresponding rates for over-reporting are $51.9\%$ and $100\%$, while the true VaR plus noise is rejected in all runs. The higher power against under-reporting by $20\%$ is consistent with its larger expected VaR score difference from the perfect-information forecast.

Comparisons between models with similar expected scores remain difficult. For example, the $20\%$ under- and over-reporting strategies are distinguished in only $0.3\%$ of the runs, although both are frequently rejected against the perfect-information standard. Thus the test can establish that two forecasts are inferior to a benchmark without being able to rank them. No false terminal colored decision is observed for any of the 30 ordered comparisons in the 1000 runs.

We also check two distinct constant forecasts with equal expected scores, taking one to be $0.8\VaR_{0.95}(L)$ and choosing the other above the true VaR to match its expected score. The forecasts are approximately $1.316$ and $2.046$. A false terminal rejection occurs in $0.1\%$ of the 1000 runs, below the stated $5\%$ error level.

\subsection{Comparative e-backtests via stationary time series data}
\label{sec:time_series}

We apply our comparative e-backtesting procedure to a stationary time series, following the settings of \cite{NZ17}, with data simulated from an AR(1)--GARCH(1,1) process:
\begin{equation*}
    L_t=\mu_t+\varepsilon_t,\quad \varepsilon_t=\sigma_tZ_t,\quad \mu_t=-0.05+0.3L_{t-1},\quad \sigma_t^2=0.01+0.1\varepsilon_{t-1}^2+0.85\sigma_{t-1}^2,\quad t\in\mathbb N,
\end{equation*}
where the innovations are iid standardized skewed-t with shape parameter $\nu=5$ and skewness parameter $\gamma=1.5$. Forecasts use rolling windows of 500 observations. After a further 500 forecasts for initialization, we assess 5000 forecasts. Thus, we index time from the first forecast after the initial estimation window, giving \(T=5500\).

For both internal and standard models, we forecast risk measures using the framework of \cite{MF00,NZ17,WWZ25}. We estimate the AR--GARCH coefficients by conditional maximum likelihood under normal, t, or skewed-t innovations with mean zero and variance one. Under the assumed location--scale model, risk forecasts are obtained by substituting the fitted quantities into $\psi(L_t\mid\mathcal F_{t-1})=\mu_t+\sigma_t\psi(Z_t)$.\footnote{VaR, ES, and expectiles are translation invariant and positively homogeneous; see \cite{A99}.} We estimate $\psi(Z_t)$ from the standardized residuals using the three approaches in \cite{NZ17}:
\begin{enumerate}
    \item[(a)] Fully parametric (FP): We compute $\psi(Z_t)$ under the fitted innovation distribution \citep{MFE15}, giving ``n-FP'', ``t-FP'' and ``st-FP''.
    \item[(b)] Filtered historical simulation (FHS): We estimate $\psi(Z_t)$ empirically from the standardized residuals \citep{C03,NZ17}, giving ``n-FHS'', ``t-FHS'' and ``st-FHS''.
    \item[(c)] Extreme value theory-based semiparametric estimation (EVT): We fit a generalized Pareto distribution to upper-tail exceedances of the standardized residuals and use the fitted tail to compute risk forecasts \citep{MF00,NZ17}, giving ``n-EVT'', ``t-EVT'' and ``st-EVT''.
\end{enumerate}
Beyond these nine models, we include optimal forecasts based on the true data-generating process, denoted ``opt'' in line with \cite{NZ17}.

For inference, $Q_{t,\theta}$ is an AR(1)--GARCH(1,1) family with standardized skewed-t innovations. Its eight unknown parameters are the intercept, AR coefficient, three GARCH coefficients, initial variance, degrees of freedom and skewness. We maintain $\alpha_0,\alpha_1,\beta_1>0$, $|\phi_1|<1$, $\alpha_1+\beta_1<1$, positive initial variance and skewness, and degrees of freedom greater than $2$. The auxiliary density $q_t$ is a separate expanding skewed-t AR--GARCH predictor, refitted daily to all preceding observations with the hybrid optimizer in the R package \texttt{rugarch}. Proposition \ref{prop:auxiliary_precision} in Section \ref{subsec:betting} gives additional sufficient conditions for asymptotic power.
All comparisons share the likelihood confidence region in \eqref{eq:likelihood_parameter_region}. Again, we adopt the overall test scheme and only use the e-three-zone approach to report colored rejections according to the score differences $\underline{\Delta}_{500,5500}$ and $\overline{\Delta}_{500,5500}$.

\begin{figure}[p]
    \centering
\includegraphics[width=0.77\linewidth]{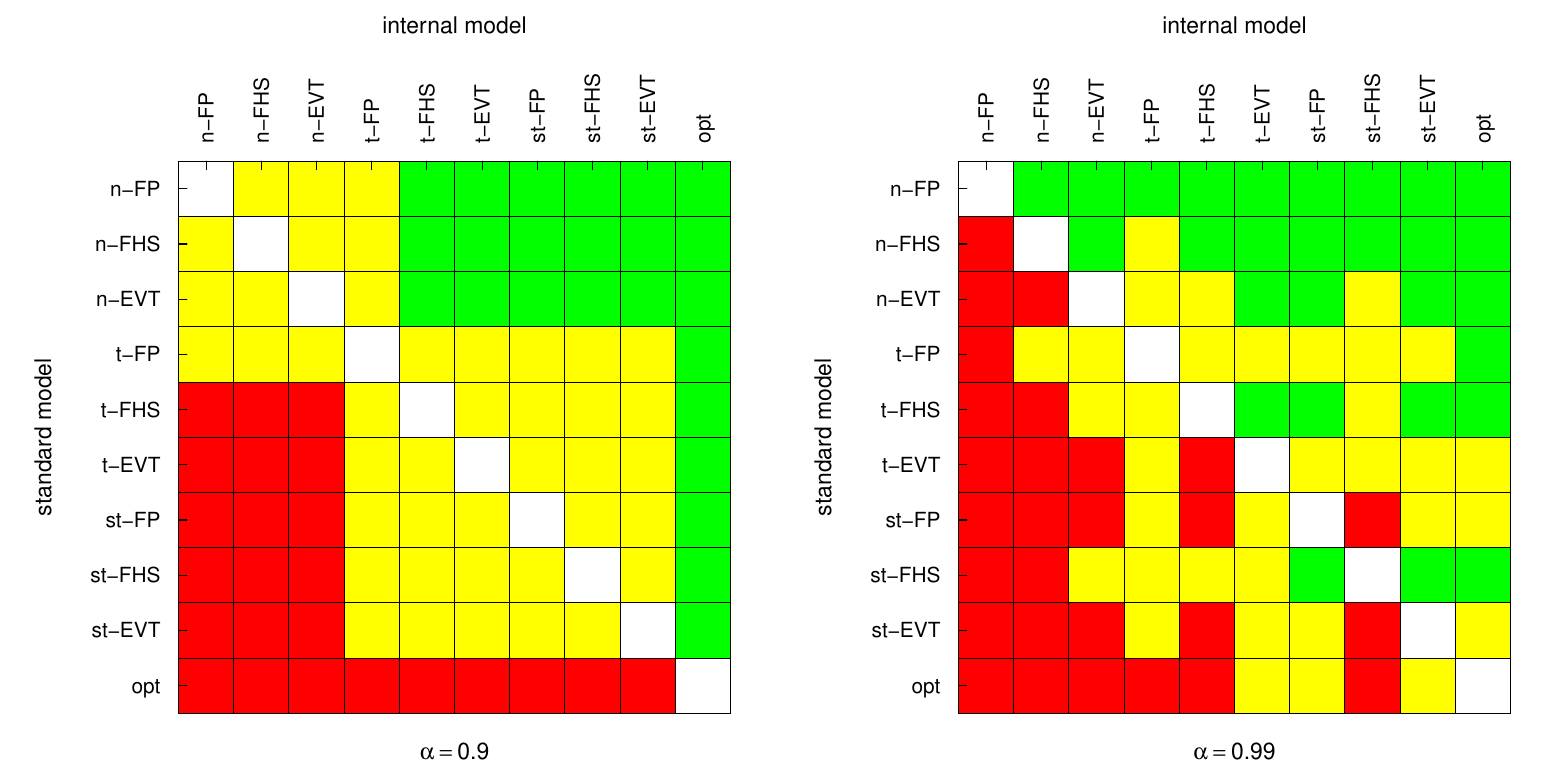}
    \caption{Overall-test heatmaps for $\VaR_p$ forecasts at levels $p=0.9$ and $0.99$ for stationary time series data. The horizontal axis represents the internal model and the vertical axis represents the standard model.}
    \label{fig:time_heat_VaR}
    \medskip
\includegraphics[width=0.77\linewidth]{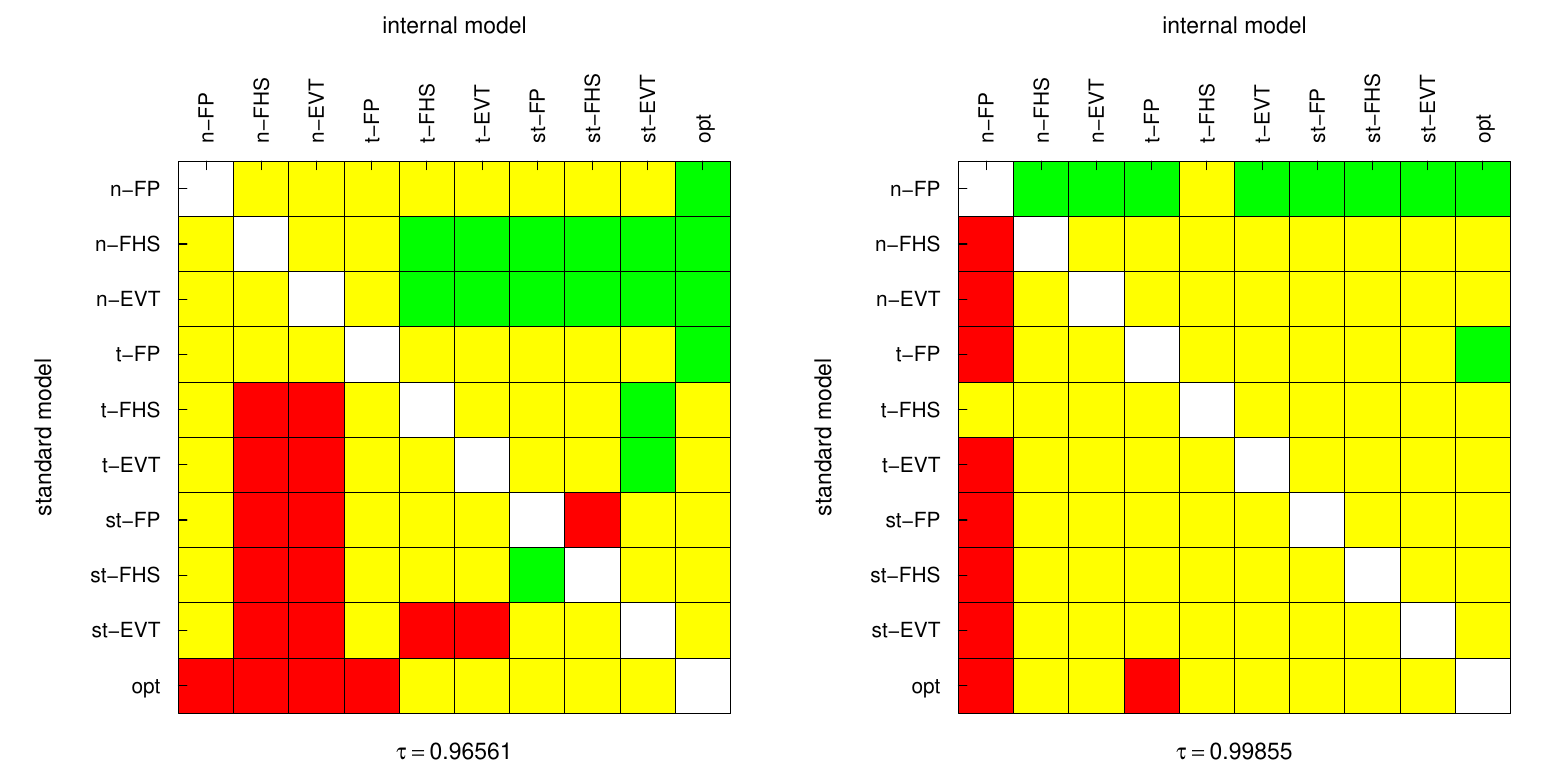}
    \caption{Overall-test heatmaps for $\ex_\tau$ forecasts at levels $\tau=0.96561$ and $0.99855$ for stationary time series data. The horizontal axis represents the internal model and the vertical axis represents the standard model.}
    \label{fig:time_heat_EXP}
    \medskip
\includegraphics[width=0.77\linewidth]{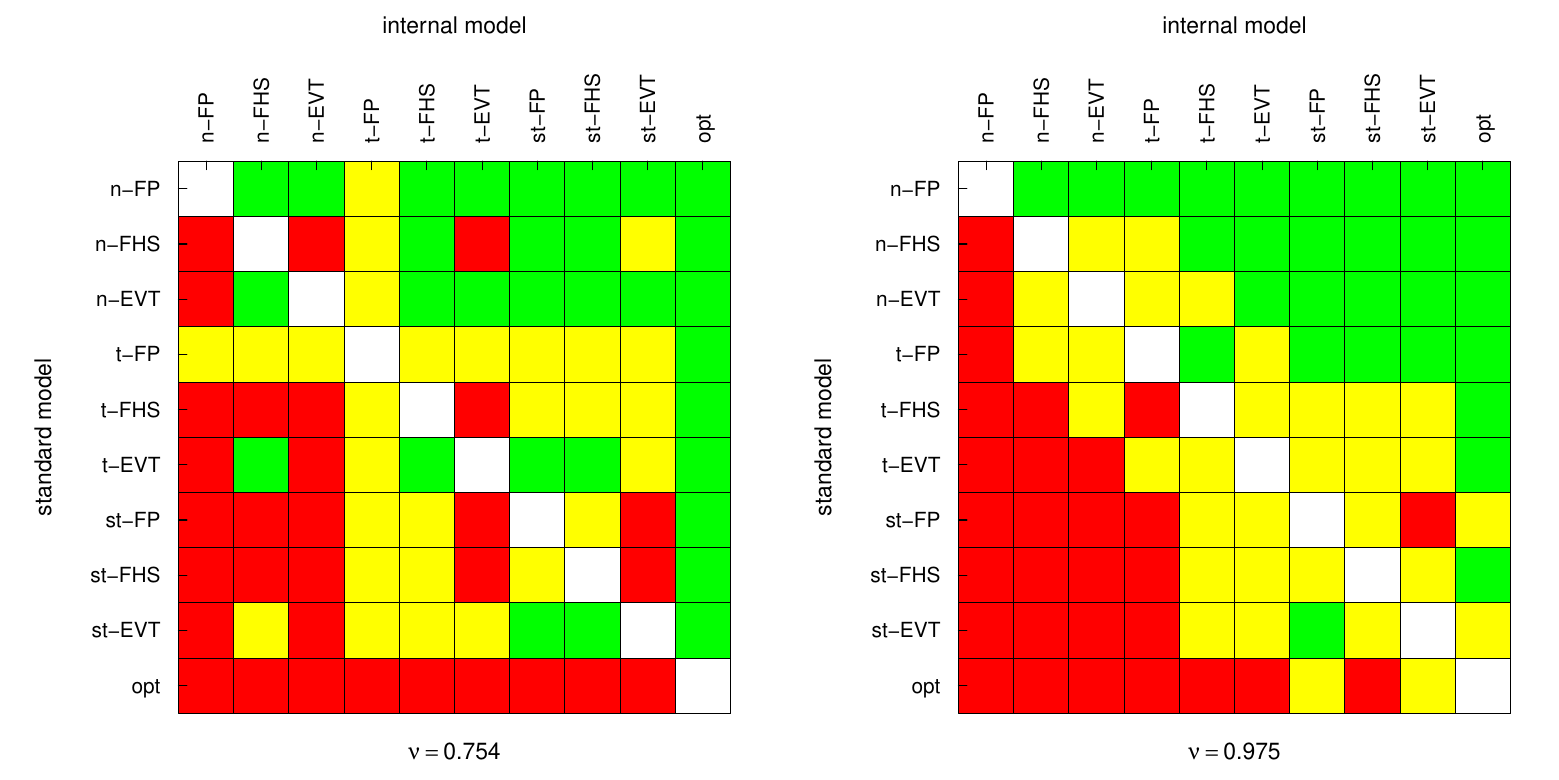}
    \caption{Overall-test heatmaps for $(\ES_\nu,\VaR_\nu)$ forecasts at levels $\nu=0.754$ and $0.975$ for stationary time series data. The horizontal axis represents the internal model and the vertical axis represents the standard model.}
    \label{fig:time_heat_ES}
\end{figure}

Figures \ref{fig:time_heat_VaR}--\ref{fig:time_heat_ES} present the overall-test heatmaps for $\VaR_p$, $\ex_\tau$, and $(\ES_\nu,\VaR_\nu)$ at typical internal risk management levels ($p=0.9$, $\tau=0.96561$, $\nu=0.754$) and higher levels ($p=0.99$, $\tau=0.99855$, $\nu=0.975$). Appendix \ref{app:sec:ts_intermediate} gives the intermediate levels $p=0.95$, $\tau=0.98761$, and $\nu=0.875$. Across the three levels, the test resolves 80 of 135 distinct pairs for VaR, 40 for expectiles and 90 for VaR--ES. All final colored directions agree with the known average conditional score differences.
The comparisons illustrate the effect of the innovation assumption. The n-FP forecasts often perform poorly when fitting the AR--GARCH process with skewed-t innovations. EVT and FHS can mitigate this misspecification by using the residual tail information, while st-FP uses the correct innovation family.

For VaR, 28 of 45 distinct pairs are resolved at $p=0.99$, compared with 24 at $p=0.9$. VaR--ES resolution is similar at its two levels, with 31 and 29 pairs resolved. Expectile comparisons become more difficult as the level increases: 17 pairs are resolved at $\tau=0.96561$, compared with only nine at $\tau=0.99855$. Eight of these nine highest-level expectile comparisons favor another model over n-FP. Appendix \ref{app:nz_replication} compares these conclusions with the fixed-time procedure of \cite{NZ17}.

\section{Real data analysis}
\label{sec:real}

In this section, we apply our comparative e-backtesting procedure to the negated percentage log-returns of the NASDAQ Composite Index from Jan 3, 2003, to May 30, 2025. We use rolling estimation windows of 500 observations and begin assessment after a further 500 forecasts, giving a reporting period from December 21, 2006, to May 30, 2025. Forecast estimation and the likelihood implementation follow Section \ref{sec:time_series}. We compare six models, n-FP, n-FHS, n-EVT, st-FP, st-FHS and st-EVT, for $\VaR_{0.99}$, $\ex_{0.99855}$ and $(\ES_{0.975},\VaR_{0.975})$.

We are particularly interested in how the comparisons change during periods such as the 2008 financial crisis and the COVID-19 pandemic. We therefore apply both the overall and resetting test schemes to all 30 ordered pairs for each risk measure.

\begin{figure}[p]
    \centering
\includegraphics[width=0.96\linewidth]{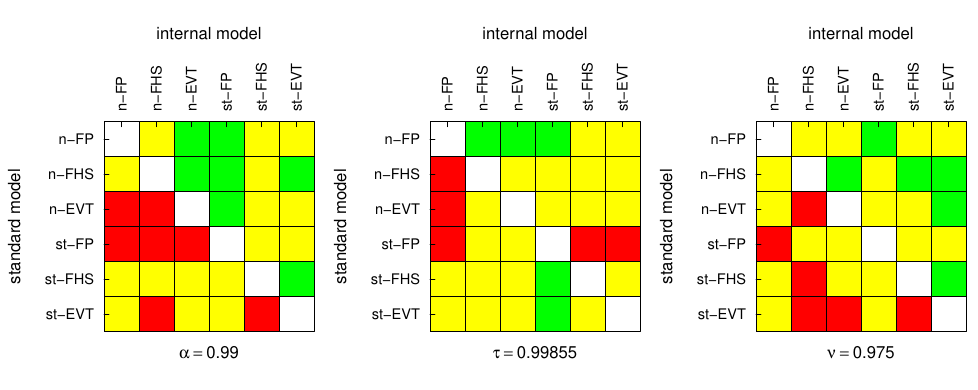}
    \caption{Overall-test heatmaps for the NASDAQ forecasts (left panel: $\VaR_{0.99}$, middle panel: $\ex_{0.99855}$, right panel: $(\ES_{0.975},\VaR_{0.975})$). The horizontal axis represents the internal model and the vertical axis represents the standard model.}
    \label{fig:realheatmap}
    \medskip
\includegraphics[width=\textwidth,height=0.62\textheight,keepaspectratio]{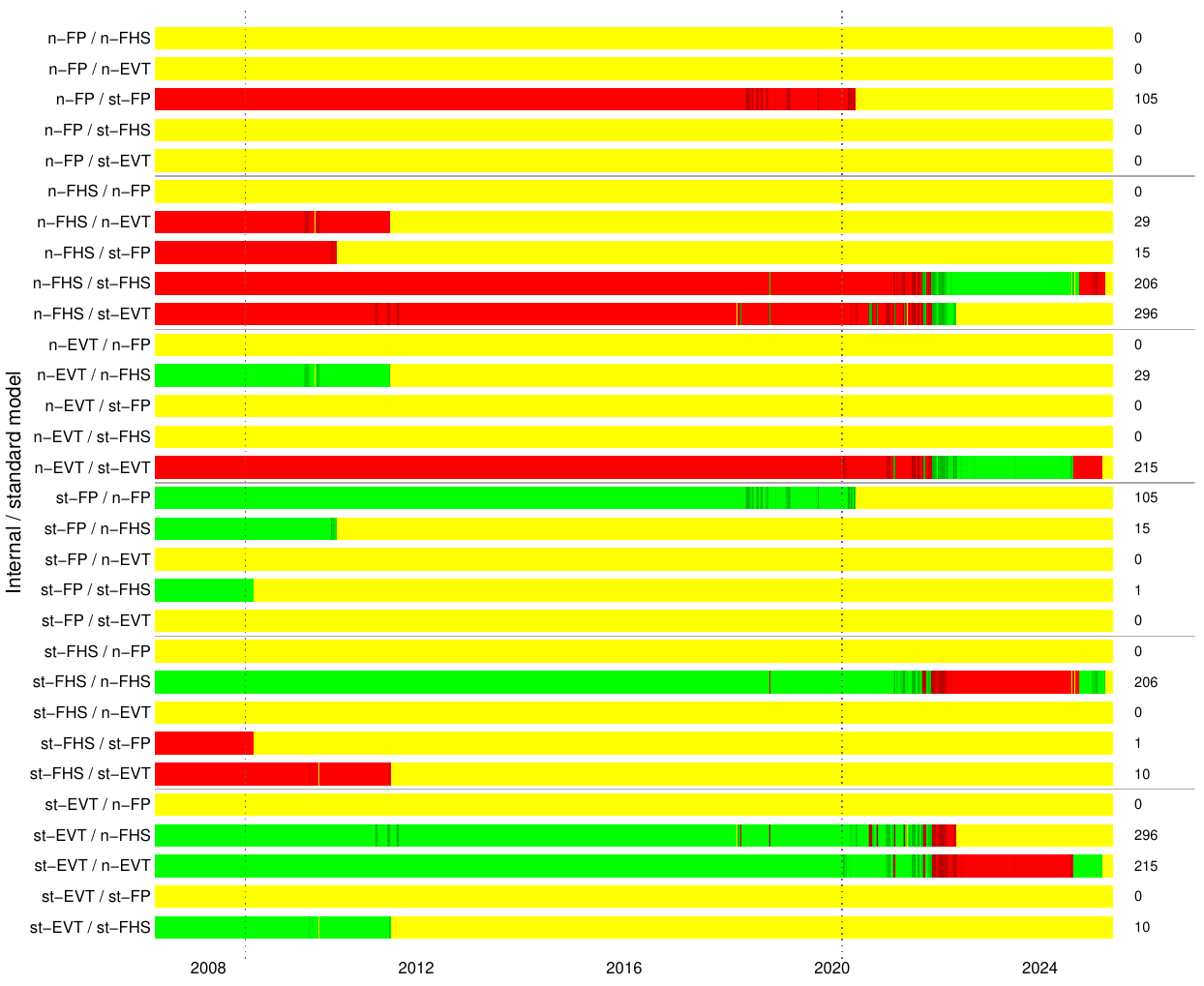}
    \caption{Resetting reports for all 30 ordered NASDAQ VaR--ES comparisons. Rejection points are marked in gray. Each row gives the internal model followed by the standard model. Numbers give separate completed report counts. Dotted lines mark September 15, 2008 and March 11, 2020.}
    \label{fig:realeprocess}
\end{figure}

Figure \ref{fig:realheatmap} presents the overall results. The test resolves seven of the 15 distinct pairs for VaR, five for expectiles and six for VaR--ES. With n-FP as the standard, there are two, three and one green comparisons, respectively. With st-FP as the standard, there are three, three and one red comparisons, respectively, and no green comparisons. The tests favor st-FP over three competitors for both VaR and expectiles. For VaR--ES, they favor st-EVT over n-FHS, n-EVT and st-FHS, while its comparison with st-FP remains yellow. The comparisons therefore favor different methods for different risk measures, without establishing a universally best model.

Figure \ref{fig:realeprocess} shows the resetting VaR--ES comparisons as colored assessment blocks. The VaR and expectile comparisons are provided in Appendix \ref{app:nasdaq_resetting}. All 30 ordered pairs are displayed, with reversed comparisons sharing report dates and interchanging red and green.
Counting each pair once, the resetting test produces 1,222 VaR reports, 336 expectile reports and 877 VaR--ES reports. Both colors occur on different assessment intervals for seven distinct VaR pairs and three distinct pairs for each of expectiles and VaR--ES. Many completed assessments cover a single forecast date, supported by the accumulated likelihood information. These totals therefore count repeated assessments of the same 15 model pairs.

The resetting results also show changes in relative average performance. For example, the 2008 financial crisis and COVID-19 pandemic period often yield conclusive rejections (e.g., st-FP vs.~st-FHS, st-FHS vs.~st-FP) or color switching (e.g., n-FHS vs.~st-FHS, n-EVT vs.~st-EVT).
In particular, comparing n-FHS with st-FHS for VaR--ES gives a red report for October 22, 2018--March 5, 2021, and a green report for May 3, 2022--August 9, 2024. Each conclusion concerns average performance over the whole assessment interval; the crisis dates in the figures provide calendar references. Resetting therefore adds information about periods of relative outperformance that is not summarized by the terminal heatmap.

\section{Conclusion}
\label{sec:conclusion}

This paper develops a comprehensive e-value-based framework for comparative backtests of risk measure forecasts. Our method is parametric and sequentially valid under the stated conditional loss model.
% Building on \citet{WWZ25}, we broaden the scope to include expectile, variantile, and other risk measures. More importantly, we establish a model-free comparative backtesting framework that enables meaningful and anytime-valid comparison between internal and standard forecasting models for financial regulatory practice.
Through simulation and real data, we demonstrate two practical approaches: (i) applying the e-backtest across the entire dataset with heatmap visualization; (ii) restarting the score assessment at chosen or random points for dynamic, sequential comparisons. For both schemes, an e-three-zone approach provides a tractable guideline for practical manipulations with sound theoretical guarantees.

Financial regulatory comparative backtesting and statistical model selection are not essentially the same problem due to their different initial purposes and practical constraints. Besides the ample studies of statistical model selection problems, we expect more future work on comparative backtests for the interests of financial regulation. Also, future research on optimizing betting strategies and incorporating the information asymmetry in comparative backtests will be of great interest.

\subsection*{Acknowledgement}
The paper has been substantially rewritten and improved from its previous version. We thank Junyi Guo, Lisa Gao, Xia Han, Gee Lee, Haiyan Liu, Liang Peng, Ruodu Wang and Fan Yang for helpful comments on the paper. ChatGPT-6.1 Sol was used in the development of technical results and in the production of code for numerical experiments.
The paper is human-written.

\subsection*{Disclosure statement}\label{disclosure-statement}

The authors claim that no conflicts of interest exist.

\subsection*{Data Availability Statement}\label{data-availability-statement}

The NASDAQ data and code for the numerical studies are available at \url{https://github.com/qwangan/Comparative-e-backtesting}.

\newpage

\setcounter{section}{0}
\renewcommand{\thesection}{\Alph{section}}

\setcounter{table}{0}
\renewcommand{\thetable}{\thesection\arabic{table}}

\setcounter{figure}{0}
\renewcommand{\thefigure}{\thesection\arabic{figure}}

\setcounter{lemma}{0}
\renewcommand{\thelemma}{\thesection.\arabic{lemma}}
\setcounter{proposition}{0}
\renewcommand{\theproposition}{\thesection.\arabic{proposition}}
\renewcommand{\theHproposition}{appendix.\thesection.\arabic{proposition}}
\setcounter{theorem}{0}
\renewcommand{\thetheorem}{\thesection.\arabic{theorem}}
\setcounter{definition}{0}
\renewcommand{\thedefinition}{\thesection.\arabic{definition}}
\setcounter{corollary}{0}
\renewcommand{\thecorollary}{\thesection.\arabic{corollary}}
\setcounter{example}{0}
\renewcommand{\theexample}{\thesection.\arabic{example}}
\setcounter{equation}{0}
\renewcommand{\theequation}{\thesection.\arabic{equation}}
\renewcommand{\theHequation}{appendix.\thesection.\arabic{equation}}
\renewcommand{\theHexample}{appendix.\thesection.\arabic{example}}

\appendix

% \section{Scoring functions for common risk measures}
% \label{app:scor}

\section{Additional technical details}
\label{app:sec:tech}

\subsection{Additional assumptions}
\label{app:sec:assum}

The following is a standard assumption for time series modeling together with continuity assumptions of the density and expected score differences.
\begin{assumption}\label{app:assum:stationary}
    $\Theta$ is compact and the losses $(L_t)_{t\in\T_+}$, forecasts $(R_t)_{t\in\T_+}$, $(R^*_t)_{t\in\T_+}$, and candidate conditional density family $(q_{t,\theta})_{t\in\T_+}$, $\theta\in\Theta$, are jointly stationary and ergodic with respect to the stationary past filtration. The log density $\log q_{1,\theta}(L_1)$ and the conditional score mean $\int_\R\{S(x,R_1)-S(x,R_1^*)\}\d Q_{1,\theta}(x)$ are continuous in $\theta$ a.s.
\end{assumption}

\subsection{Student and skewed-t innovation densities}
\label{app:sec:innovation_densities}

The following densities have mean zero and variance one. For $\nu>2$, the standardized Student density is
\begin{equation*}
    f_\nu(z)=\frac{\Gamma((\nu+1)/2)}{\sqrt{\pi(\nu-2)}\,\Gamma(\nu/2)}\left(1+\frac{z^2}{\nu-2}\right)^{-(\nu+1)/2},\qquad z\in\R.
\end{equation*}
For $\nu>2$ and $\gamma>0$, the standardized skewed-t density of \cite{FS98} is
\begin{equation}\label{app:eq:skew_student_density}
    f_{\nu,\gamma}(z)=\frac{2s}{\gamma+\gamma^{-1}}\left\{f_\nu(\gamma(sz+m))\id_{\{sz+m<0\}}+f_\nu((sz+m)/\gamma)\id_{\{sz+m\ge0\}}\right\},\qquad z\in\R,
\end{equation}
where $s>0$ and
\[
    m=\frac{2\sqrt{\nu-2}\,\Gamma((\nu+1)/2)}{\sqrt\pi(\nu-1)\Gamma(\nu/2)}(\gamma-\gamma^{-1}),\qquad s^2=\gamma^2+\gamma^{-2}-1-m^2.
\]
Setting $\gamma=1$ gives $f_{\nu,1}=f_\nu$.

\subsection{AR--GARCH specification with an unknown initial variance}
\label{app:sec:argarch}

Consider a correctly specified stationary AR(1)--GARCH(1,1) model with iid standardized Student or skewed-t innovations. Let $\Theta$ consist of the structural parameters satisfying \eqref{eq:argarch_power_bounds} and the initial conditional variance $\sigma_1^2\ge\alpha_0$, with no upper bound. This lower bound follows from the stationary GARCH recursion. Use $q_t=q_{t,\widehat\theta_{t-1}}$, estimating all coordinates of $\theta$, including $\sigma_1^2$, jointly. Fix $\lambda_t=\lambda\in(0,1)$ and replace \eqref{eq:likelihood_parameter_region} by
\begin{equation}\label{eq:likelihood_nested_region}
    \mathcal C_t=\{\theta\in\Theta:\max_{0\le u\le t}E_u(\theta)<1/\alpha\}.
\end{equation}
Theorem~\ref{thm:adjusted_eprocess} and Proposition~\ref{prop:reset_fdr} remain valid for this running intersection.

\begin{proposition}\label{prop:argarch_power}
    Suppose the forecasts are jointly stationary and ergodic with the losses and satisfy the first condition in \eqref{eq:raw_location_scale_precision}. For joint fits satisfying \eqref{eq:auxiliary_fit} over $\Theta$, the conclusions of Proposition~\ref{prop:auxiliary_precision} hold for \eqref{eq:likelihood_nested_region}. An exact joint conditional MLE exists and satisfies \eqref{eq:auxiliary_fit}.
\end{proposition}

\subsection{Constructing approximate confidence regions}
\label{app:sec:confidence_computation}

We show the detailed procedures we follow to construct approximate confidence regions $\widehat{\mathcal C}_{t-1}$ used to calculate the bounds of the average score differences in \eqref{eq:auxiliary_projection}.

\emph{IID normal losses.} For $\theta=(\mu,\sigma)$, construct an enclosing polygon as follows.

Let $n=t-l-1$. Report yellow if $n<3$; otherwise, let $\bar L$ and $\hat\sigma>0$ be the sample mean and standard deviation of $L_{l+1},\ldots,L_{t-1}$, using divisor $n$. Since $\log\{(1+e^x)/2\}\ge x/2$, every $\theta\in\mathcal C_{t-1}$ satisfies
\[
    \sum_{u=l+1}^{t-1}\log q_{u,\theta}(L_u)>\sum_{u=l+1}^{t-1}\log q_u(L_u)-2\log20.
\]
Set
\[
    d=-\log(\sqrt{2\pi}\hat\sigma)-\frac12-\frac1n\sum_{u=l+1}^{t-1}\log q_u(L_u)+\frac{2\log20}{n}.
\]
If $d\le0$, the region is empty. Otherwise, the normal likelihood gives $\mu_-\le\mu\le\mu_+$ and $\sigma_-\le\sigma\le\sigma_+$, where $\mu_\pm=\bar L\pm\hat\sigma\sqrt{e^{2d}-1}$
and $\sigma_-<\hat\sigma<\sigma_+$ are the two roots of
\[
    \log(\sigma/\hat\sigma)+\hat\sigma^2/(2\sigma^2)-1/2=d,
\]
calculated by bisection. These bounds give a rectangle containing $\mathcal C_{t-1}$.

We narrow the enclosing rectangle to obtain tighter bounds in \eqref{eq:auxiliary_projection}. Set $(a,b)=(\mu/\sigma,1/\sigma)$ and $f(a,b)=\log E_{t-1}(a/b,1/b)$, which is convex. The enclosing rectangle in these coordinates has $1/\sigma_+\le b\le1/\sigma_-$, with $a$ between the smallest and largest of the four corner ratios $\mu_-/\sigma_-$, $\mu_-/\sigma_+$, $\mu_+/\sigma_-$ and $\mu_+/\sigma_+$. Choose 40 points \((a_k,b_k)\) satisfying \(f(a_k,b_k)=\log20\). By convexity, each point gives the following half-plane containing the confidence region:
\[
    f(a_k,b_k)+f_a(a_k,b_k)(a-a_k)+f_b(a_k,b_k)(b-b_k)\le\log20.
\]
Intersect these half-planes with this rectangle and map the polygon back to $(\mu,\sigma)$ to obtain $\widehat{\mathcal C}_{t-1}\supseteq\mathcal C_{t-1}$. Triangulate the polygon and bound the average score difference on each triangle, allowing at most 2048 subdivisions. Report yellow if neither sign is established.

\emph{AR--GARCH losses.} Let $\Theta_{\mathrm{box}}$ be the subset of $\Theta$ satisfying the following numerical bounds.
\begin{center}
\begin{tabular}{lrr}
\toprule
Coordinate & Lower bound & Upper bound\\
\midrule
$\phi_0/(1-\phi_1)$ & $-100$ & $100$\\
$\operatorname{arctanh}(\phi_1)$ & $-8$ & $8$\\
$\log\alpha_0$ & $-40$ & $12$\\
$\log\{\alpha_1/(1-\alpha_1-\beta_1)\}$ & $-28$ & $28$\\
$\log\{\beta_1/(1-\alpha_1-\beta_1)\}$ & $-28$ & $28$\\
$\log\sigma_0^2$ & $-40$ & $30$\\
$\log(\nu-2)$ & $-16$ & $18$\\
$\log\gamma$ & $-8$ & $8$\\
\bottomrule
\end{tabular}
\end{center}
We use the approximate region
\[
    \widehat{\mathcal C}_{t-1}=\{\theta\in\Theta_{\mathrm{box}}:\log E_{t-1}(\theta)\le\log20\}.
\]
We minimize and maximize $\Delta_{s,t}(\theta)$ numerically over $\widehat{\mathcal C}_{t-1}$ using SLSQP in \texttt{nloptr}.
The search box and local optimization give numerical approximations to \eqref{eq:auxiliary_projection}.

\section{Additional examples}
\label{app:sec:ex}

\subsection{Identification functions for common risk measures}
\label{app:sec:iden}

The following example shows identification functions for common risk measures.
Take $p\in(0,1)$, with finite first moments for the mean, ES and expectile, and finite second moments for the variance and variantile. The VaR and VaR--ES entries assume continuous loss distributions. The following bounds give admissible choices of $h$ in \eqref{eq:naive_e1}. When $a>0$ is required, the corresponding risk-measure component is assumed positive.

\begin{example}\label{app:ex:1}
\begin{enumerate}
    \item (The mean) For non-negative losses, use
    \[
        I(x,a)=\frac xa-1,\qquad x\ge0,\ a>0.
    \]
    Since $I(x,a)\ge-1$, take $0\le h\le1$.

    \item (The mean and the variance) For $(\Var,\E)$, use
    \[
        I(x,a,b)=\left(x-b,\frac{(x-b)^2}{a}-1\right)^\top,
        \qquad x,b\in\R,\ a>0.
    \]
    Since $I_2(x,a,b)\ge-1$, take $h=(0,h_2)^\top$ with $0\le h_2\le1$.

    \item (The Value-at-Risk) For $\VaR_p$, use
    \[
        I(x,a)=\frac{\id_{\{x>a\}}}{1-p}-1,\qquad x,a\in\R.
    \]
    Since $-1\le I(x,a)\le p/(1-p)$, take $-(1-p)/p\le h\le1$.

    \item (The Value-at-Risk and the Expected Shortfall) For $(\ES_p,\VaR_p)$, use
    \[
        I(x,a,b)=
        \left(\frac{\id_{\{x>b\}}}{1-p}-1,
        \frac{(x-b)_+}{(1-p)(a-b)}-1\right)^\top,
        \qquad x,b\in\R,\ a>b.
    \]
    Both components are at least $-1$, so take $h=(h_1,h_2)^\top$ with $h_1,h_2\ge0$ and $h_1+h_2\le1$.

    \item (The expectile) For non-negative losses and $\ex_p$ \citep{NP87}, use
    \[
        I(x,a)=|1-p-\id_{\{x>a\}}|\left(\frac xa-1\right),
        \qquad x\ge0,\ a>0.
    \]
    Since $I(x,a)\ge-(1-p)$, take $0\le h\le1/(1-p)$.

    \item (The expectile and the variantile) For $(\var_p,\ex_p)$ \citep{FK20}, use
    \[
        \begin{aligned}
            I(x,a,b)&=\left(|1-p-\id_{\{x>b\}}|(x-b),
            \frac{2\{p(x-b)_+^2+(1-p)(x-b)_-^2\}}{a}-1\right)^\top,\\
            &\qquad x,b\in\R,\ a>0.
        \end{aligned}
    \]
    Since $I_2(x,a,b)\ge-1$, take $h=(0,h_2)^\top$ with $0\le h_2\le1$.
\end{enumerate}
\end{example}

\subsection{Scoring functions for common risk measures}
\label{app:sec:score1}

The following examples give scoring functions for common risk measures. Take distributions with finite second moments for the mean, variance and squared expectile scores, and finite first moments for the other scores. For VaR and ES, use continuous loss distributions. For the logarithmic VaR and expectile scores, take positive target values; for the VaR--ES scores, take positive ES.

\begin{example}\label{ex:2}
\begin{enumerate}
    \item (The mean) The mean has the strictly consistent scoring function
    \begin{equation*}
        S_{\E}(x,a)=(x-a)^2,\qquad x,a\in\R.
    \end{equation*}

    \item (The mean and the variance) For positive variance, the pair $(\Var,\E)$ has the strictly consistent scoring function
    \begin{equation*}
        S_{\Var,\E}(x,a,b)=b(b-2x)+(a+b^2)(a+b^2-2x^2),
        \qquad x,b\in\R,\ a>0.
    \end{equation*}

    \item (The Value-at-Risk) For $p\in(0,1)$, the VaR score is
    \begin{equation*}
        S_{\VaR}(x,a)=(1-p)a+(x-a)_+,\qquad x,a\in\R.
    \end{equation*}

    \item (The Value-at-Risk and the Expected Shortfall) The pair $(\ES_p,\VaR_p)$ has the strictly consistent scoring function
    \begin{equation}\label{eq:scoreVaRES}
        S_{\ES,\VaR}(x,a,b)=\frac{(x-b)_+}{2\sqrt a}
        +(1-p)\frac{a+b}{2\sqrt a},\qquad x,b\in\R,\ a>0,\ b\le a.
    \end{equation}

    \item (The expectile) For $p\in[1/2,1)$, the expectile score is
    \begin{equation*}
        S_{\ex}(x,a)=(2p-1)(x-a)_+^2+(1-p)a(a-2x),\qquad x,a\in\R.
    \end{equation*}
\end{enumerate}
\end{example}

For distinct forecasts, the mean, $(\Var,\E)$ and expectile score differences are unbounded. The VaR--ES difference is unbounded when the ES forecasts differ; the VaR difference is bounded.

\subsection{Scores with \texorpdfstring{$0$}{0}-homogeneous differences}
\label{app:sec:score0}

The following scores have $0$-homogeneous differences. In indicator-weighted logarithmic terms, the term is defined as zero when the indicator is zero.

\begin{example}\label{app:ex:0}
\begin{enumerate}
    \item (The Value-at-Risk) For $a>0$, the strictly consistent score is
    \[
        S_{\VaR}(x,a)=(1-p)\log a+\id_{\{x>a\}}\log\frac xa,\qquad x\in\R.
    \]

    \item (The Value-at-Risk and the Expected Shortfall) The strictly consistent score is
    \[
        S_{\ES,\VaR}(x,a,b)=\frac{(x-b)_+}{a}
        +(1-p)\left(\frac ba-1+\log a\right),
        \qquad x,b\in\R,\ a>0,\ b\le a.
    \]

    \item (The expectile) For $p\in[1/2,1)$ and $a>0$, the strictly consistent score is
    \[
        \begin{aligned}
            S_{\ex}(x,a)={}&(1-2p)\id_{\{x>a\}}
            \left(\log\frac xa+1-\frac xa\right)\\
            &+(1-p)\left(\log a-1+\frac xa\right),\qquad x\in\R.
        \end{aligned}
    \]
\end{enumerate}
\end{example}

The VaR score difference is bounded. The expectile score difference is unbounded for distinct forecasts, and the VaR--ES score difference is unbounded when the ES forecasts differ.

\section{Proofs for all results}
\label{app:sec:proofs}

\begin{proof}[Proof of Theorem~\ref{thm:adjusted_eprocess}]
    Under \eqref{eq:auxiliary_correction}, predictability and normalization give
    \[
        \E\!\left[1-\lambda_t+\lambda_t
           \frac{q_t(L_t)}{q_{t,\theta_0}(L_t)}
           \middle|\mathcal F_{t-1}\right]
        =1-\lambda_t+\lambda_t\int_\R q_t(x)\d x=1.
    \]
    Thus $E_t(\theta_0)$ is a non-negative martingale with $E_0(\theta_0)=1$. Ville's inequality gives \eqref{eq:auxiliary_cs}.
    On the event $\{\theta_0\in\mathcal C_t\text{ for all }t\in\T\}$, we have $\theta_0\in\mathcal C_{t-1}$ for every $t\in\T_+$. Equations \eqref{eq:auxiliary_correction} and \eqref{eq:auxiliary_projection} give
    $\underline\Delta_{s,t}\le\Delta_{s,t}(\theta_0)=\Delta_{s,t}\le\overline\Delta_{s,t}$
    simultaneously for all $0\le s<t$. Neither event in
    \eqref{eq:directional_validity} can then occur, proving \eqref{eq:directional_validity}.
\end{proof}

\begin{proof}[Proof of Proposition~\ref{prop:adjusted_power}]
    Under the red alternative $\liminf_{t\to\infty}\Delta_t\ge\varepsilon$ a.s., as $t\to\infty$, we have $\p(\Delta_t\le\varepsilon/2)\to0$ and
    \[
    \begin{aligned}
        \p(\underline\Delta_{0,t}\le0)
        &\le \p(\Delta_t\le\varepsilon/2)
        +\p(|\underline\Delta_{0,t}-\Delta_t|\ge\varepsilon/2)
        \to0.
    \end{aligned}
    \]
    Reversing the signs and replacing $\underline{\Delta}_{0,t}$ by $\overline{\Delta}_{0,t}$ give the other convergence. If both convergences in \eqref{eq:prop1_conv} hold a.s., then
    \[
        \liminf_{t\to\infty}\underline\Delta_{0,t}
          =\liminf_{t\to\infty}\Delta_t,
        \qquad
        \limsup_{t\to\infty}\overline\Delta_{0,t}
          =\limsup_{t\to\infty}\Delta_t
        \quad\text{a.s.},
    \]
    which gives the two probability-one assertions.
\end{proof}

\begin{proof}[Proof of Proposition~\ref{prop:reset_fdr}]
    By \eqref{eq:auxiliary_correction} and \eqref{eq:auxiliary_projection}, for all $s,t\in\T$ with $s<t$ and $\theta_0\in\mathcal C_{t-1}$,
    \[
        \underline\Delta_{s,t}\le\Delta_{s,t}(\theta_0)=\Delta_{s,t}\le\overline\Delta_{s,t}~~~\mbox{a.s.}
    \]
    Substituting $(s,t)=(\tau_{k-1},\tau_k)$ on $\{\tau_k<\infty\}$ gives
    \[
        \{\tau_k<\infty,\ k\text{-th report is false}\}\subseteq\{\tau_k<\infty,\theta_0\notin\mathcal C_{\tau_k-1}\}\subseteq\{\exists\,u\in\T:\theta_0\notin\mathcal C_u\},~~k\in\N,~~~\mbox{a.s.}
    \]
    Since $0\le V_t\le N_t$, for every $\T_+$-valued stopping time $\tau$,
    \[
        \frac{V_\tau}{N_\tau\vee1}\le\sup_{t\in\T}\frac{V_t}{N_t\vee1}\le\id_{\{\exists\,u\in\T:\theta_0\notin\mathcal C_u\}}\quad\text{a.s.}
    \]
    Taking expectations and applying \eqref{eq:auxiliary_cs} yields
    \[
        \mathrm{FDR}_\tau\le\mathrm{FDR}_\infty\le\p(\exists\,u\in\T:\theta_0\notin\mathcal C_u)\le\alpha.\qedhere
    \]
\end{proof}

\begin{proof}[Proof of Proposition~\ref{prop:auxiliary_precision}]
    Write $\Theta_0=\{\theta\in\Theta:\E\operatorname{KL}(Q_{1,\theta_0}\Vert Q_{1,\theta})=0\}$,\footnote{The Kullback--Leibler divergence is $\operatorname{KL}(Q_{u,\theta_0}\Vert Q_{u,\theta})=\int_\R\log\{q_{u,\theta_0}(x)/q_{u,\theta}(x)\}\d Q_{u,\theta_0}(x)$.} so $\theta_0\in\Theta_0$, and use $\inf\varnothing=\infty$. Assumption~\ref{app:assum:stationary} and \eqref{eq:auxiliary_envelope} give
    \[
        \sup_{\theta\in\Theta}\left|\frac1t\sum_{u=1}^{t-1}\log\frac{q_{u,\theta_0}(L_u)}{q_{u,\theta}(L_u)}-\E\operatorname{KL}(Q_{1,\theta_0}\Vert Q_{1,\theta})\right|\to0\quad\text{a.s.}
    \]
    The expected KL divergence is continuous in $\theta$. By compactness, for each $\delta>0$,
    \[
        \inf_{\theta\in\Theta:\,\operatorname{dist}(\theta,\Theta_0)\ge\delta}\E\operatorname{KL}(Q_{1,\theta_0}\Vert Q_{1,\theta})>0.
    \]
    Together with \eqref{eq:auxiliary_fit}, this implies $\operatorname{dist}(\widehat\theta_t,\Theta_0)\to0$ a.s. Nonnegativity of KL, continuity and separability give $\sup_{\theta\in\Theta_0}|\log\{q_{u,\theta_0}(L_u)/q_{u,\theta}(L_u)\}|=0$ a.s.~for all $u$. Hence, by the ergodic theorem and dominated convergence,
    \[
        \limsup_{t\to\infty}\frac1t\sum_{u=1}^{t-1}\left|\log\frac{q_{u,\theta_0}(L_u)}{q_u(L_u)}\right|\le\E\!\left[\sup_{\theta\in\Theta:\,\operatorname{dist}(\theta,\Theta_0)\le\delta}\left|\log\frac{q_{1,\theta_0}(L_1)}{q_{1,\theta}(L_1)}\right|\right]\to0\quad(\delta\downarrow0),\quad\text{a.s.}
    \]

    The second bound in \eqref{eq:auxiliary_envelope} makes $\E[\Delta_{0,1}(\theta)-\Delta_1]$ finite and continuous in $\theta$. The uniform score law is
    \[
        \sup_{\theta\in\Theta}\left|\Delta_{0,t}(\theta)-\Delta_t-\E[\Delta_{0,1}(\theta)-\Delta_1]\right|\to0\quad\text{a.s.}
    \]
    Since $\E[\Delta_{0,1}(\theta)-\Delta_1]=0$ on $\Theta_0$, compactness gives, for every $\varepsilon>0$,
    \[
        \inf_{\theta\in\Theta:\,|\E[\Delta_{0,1}(\theta)-\Delta_1]|\ge\varepsilon/2}\E\operatorname{KL}(Q_{1,\theta_0}\Vert Q_{1,\theta})>0.
    \]
    Moreover, $\log(1-\lambda+\lambda r)\ge\lambda\log r$ gives
    \[
        \frac{\log E_{t-1}(\theta)}t\ge\frac\lambda t\sum_{u=1}^{t-1}\log\frac{q_u(L_u)}{q_{u,\theta_0}(L_u)}+\frac\lambda t\sum_{u=1}^{t-1}\log\frac{q_{u,\theta_0}(L_u)}{q_{u,\theta}(L_u)}.
    \]
    Combining these bounds yields
    \[
        \liminf_{t\to\infty}\inf_{\theta\in\Theta:\,|\Delta_{0,t}(\theta)-\Delta_t|\ge\varepsilon}\frac{\log E_{t-1}(\theta)}t>0\quad\text{a.s.}
    \]
    Thus, for each $\varepsilon>0$, eventually a.s.,
    \[
        \mathcal C_{t-1}\ne\varnothing\quad\implies\quad\max\{|\underline\Delta_{0,t}-\Delta_t|,|\overline\Delta_{0,t}-\Delta_t|\}\le\varepsilon.
    \]
    Apply this to $\varepsilon=1,1/2,\ldots$. The coverage event in \eqref{eq:auxiliary_cs} guarantees nonemptiness and has probability at least $1-\alpha$, proving both assertions.

    % The uniform score law also gives $\Delta_{0,t}(\widehat\theta_{t-1})-\Delta_t\to0$ a.s. Thus replacing an empty $\mathcal C_{t-1}$ by $\{\widehat\theta_{t-1}\}$ yields consistent, predictable endpoints; $\{\mathcal C_{t-1}=\varnothing\}\subseteq\{\theta_0\notin\mathcal C_{t-1}\}$ shows that simultaneous coverage is unchanged.

    For the location--scale conditions, \eqref{eq:raw_location_scale_precision} implies
    \[
        \sup_{\theta\in\Theta}\int_{|x|>M}|x|\d Q_{1,\theta}(x)\le\frac1M\sup_{\theta\in\Theta}\int_\R x^2\d Q_{1,\theta}(x)\to0\quad(M\to\infty),\quad\text{a.s.}
    \]
    Together with weak continuity, this proves score continuity. Furthermore,
    \[
    \begin{aligned}
        &\sup_{\theta\in\Theta}\int_\R|S(x,R_1)-S(x,R_1^*)-S(0,R_1)+S(0,R_1^*)|\d Q_{1,\theta}(x)\\
        &\qquad\le\operatorname{Lip}\{S(\cdot,R_1)-S(\cdot,R_1^*)\}\sup_{\theta\in\Theta}\{|\mu_1(\theta)|+\sigma_1(\theta)\E|Z_\theta|\}.
    \end{aligned}
    \]
    Cauchy--Schwarz and \eqref{eq:raw_location_scale_precision} make the right side integrable, proving the second bound in \eqref{eq:auxiliary_envelope}.
\end{proof}

\begin{proof}[Proof of Proposition~\ref{prop:argarch_power}]
    We first verify the conditions of Proposition~\ref{prop:auxiliary_precision} for stationary conditional densities. Use $f_{\nu,\gamma}$ from \eqref{app:eq:skew_student_density}, with $\gamma=1$ for Student innovations. The conditional density is
    \[
        q_{u,\theta}(x)=\frac1{\sigma_u(\theta)}f_{\nu,\gamma}\!\left(\frac{x-\mu_u(\theta)}{\sigma_u(\theta)}\right),\qquad \mu_u(\theta)=\phi_0+\phi_1L_{u-1}.
    \]
    On a two-sided stationary extension of the losses, write
    \[
        \sigma_u^{\circ2}(\theta)=\frac{\alpha_0}{1-\beta_1}+\alpha_1\sum_{j=0}^{\infty}\beta_1^j(L_{u-1-j}-\phi_0-\phi_1L_{u-2-j})^2.
    \]
    A superscript $\circ$ on $q_{u,\theta}$ or $\Delta_{0,t}(\theta)$ means that $\sigma_u^{\circ}$ replaces the initialized filter $\sigma_u$. These stationary quantities do not depend on $\sigma_1^2(\theta)$, so their parameter set can be restricted to the compact set $\{\theta:\sigma_1^2(\theta)=B\}$. The parameter bounds and unit innovation variance give
    \[
        \E L_0^2\le\frac{B^2}{\varepsilon^2}+\frac{B}{\varepsilon^2(2-\varepsilon)}<\infty.
    \]
    Each tail of the variance series satisfies
    \[
        \sup_{\theta\in\Theta}\alpha_1\sum_{j=n}^{\infty}\beta_1^j(L_{u-1-j}-\phi_0-\phi_1L_{u-2-j})^2\le3\sum_{j=n}^{\infty}(1-\varepsilon)^j(B^2+L_{u-1-j}^2+L_{u-2-j}^2),
    \]
    and the series on the right at $n=0$ has expectation $3(B^2+2\E L_0^2)/\varepsilon<\infty$. Hence the filters are continuous in the structural parameters and $\E\sup_\theta\{\mu_1(\theta)^2+\sigma_1^{\circ2}(\theta)\}<\infty$. The innovation density formulas give
    \[
        C^{-1}(1+z^2)^{-(B+1)/2}\le f_{\nu,\gamma}(z)\le C,\qquad |z\,\partial_z\log f_{\nu,\gamma}(z)|\le C,
    \]
    uniformly for $2+\varepsilon\le\nu\le B$ and $\varepsilon\le\gamma\le B$, for some constant $C>0$. Since $\sigma_1^{\circ2}\ge\varepsilon$,
    \[
        \sup_\theta|\log q_{1,\theta}^{\circ}(L_1)|\le C\{1+L_1^2+L_0^2+\sup_\theta\sigma_1^{\circ2}(\theta)\},
    \]
    which is integrable. The innovation laws are weakly continuous by Scheff\'e's lemma and have second moment one.

    We next show that the fitted initial variance satisfies $\log\{1+\sigma_1^2(\widehat\theta_t)\}=o(t)$. The GARCH recursion gives
    \[
        \sigma_u^2(\theta)-\sigma_u^{\circ2}(\theta)=\beta_1^{u-1}\{\sigma_1^2(\theta)-\sigma_1^{\circ2}(\theta)\}.
    \]
    The bound on the derivative of the log density and the variance lower bound $\varepsilon$ imply
    \[
        \left|\log\frac{q_{u,\theta}(L_u)}{q_{u,\theta}^{\circ}(L_u)}\right|\le C(1-\varepsilon)^{u-1}\{\sigma_1^2(\theta)+\sigma_1^{\circ2}(\theta)\}.
    \]
    In particular, the sum over $u$ of the suprema on $\{\theta:\sigma_1^2(\theta)\le M\}$ is finite a.s. for each fixed $M$.

    For comparison, let $\theta'$ have the same structural coordinates as $\theta$ and initial variance $B\ge\alpha_0$. Then
    \[
        \sigma_u^2(\theta)=\sigma_u^2(\theta')+\beta_1^{u-1}\{\sigma_1^2(\theta)-B\}.
    \]
    Since $\sum_{u\ge1}\p(|L_u|>u)\le\E L_0^2\sum_{u\ge1}u^{-2}<\infty$, Borel--Cantelli and the variance recursion give
    \[
        \sup_{\theta:\,\sigma_1^2(\theta)=B}\{\mu_u(\theta)^2+\sigma_u^2(\theta)\}\le C(u+1)^2\quad\text{a.s.}
    \]
    Here and below, pathwise constants $C$ may depend on the loss path but are finite a.s. and independent of $u$, $t$ and $\theta$. For $\sigma_1^2(\theta)\ge B$, the density bounds give
    \[
        \log\frac{q_{u,\theta}(L_u)}{q_{u,\theta'}(L_u)}\le-\frac12\log\frac{\sigma_u^2(\theta)}{\sigma_u^2(\theta')}+C\log(u+1),~~~\left|\log\frac{q_{u,\theta}(L_u)}{q_{u,\theta'}(L_u)}\right|\le C\{\sigma_1^2(\theta)-B\}\beta_1^{u-1}.
    \]
    If $\beta_1>0$ and $\sigma_1^2(\theta)>B+1$, set
    \[
        n=\min\{t,1+\lfloor\log(\sigma_1^2(\theta)-B)/|\log\beta_1|\rfloor\}.
    \]
    For $u\le n$, use $\sigma_u^2(\theta)\ge\{\sigma_1^2(\theta)-B\}\beta_1^{u-1}$ and $(n-1)|\log\beta_1|\le\log\{\sigma_1^2(\theta)-B\}$. If $n<t$, the remaining geometric sum satisfies $\sum_{u>n}\{\sigma_1^2(\theta)-B\}\beta_1^{u-1}\le1/\varepsilon$. Summing the preceding two density bounds therefore yields
    \[
        \sum_{u=1}^t\log\frac{q_{u,\theta}(L_u)}{q_{u,\theta'}(L_u)}\le-\frac n4\log\{\sigma_1^2(\theta)-B\}+Cn\log(n+1)+C.
    \]
    Since $n\le1+\log\{\sigma_1^2(\theta)-B\}/|\log(1-\varepsilon)|$, increasing $C$ to include bounded initial variances gives
    \[
        \sum_{u=1}^t\log\frac{q_{u,\theta}(L_u)}{q_{u,\theta'}(L_u)}\le C-\frac1{16}\log\{1+\sigma_1^2(\theta)\},\qquad t\ge1.
    \]
    For $\beta_1=0$, this follows from the first likelihood term. Choose a finite integer $M\ge B$ with $\log(1+M)>16C$. Every candidate with $\sigma_1^2(\theta)>M$ has smaller likelihood than $\theta'$. Continuity on the compact set $\{\theta:\sigma_1^2(\theta)\le M\}$ therefore gives an exact joint MLE and zero likelihood gap.

    For fits satisfying \eqref{eq:auxiliary_fit}, the same comparison gives $\log\{1+\sigma_1^2(\widehat\theta_t)\}=o(t)$ a.s. Uniform summability of the initialization errors on $\{\sigma_1^2(\theta)=B\}$ also gives
    \[
        \sup_{\theta:\,\sigma_1^2(\theta)=B}\sum_{u=1}^t\log\frac{q_{u,\theta}^{\circ}(L_u)}{q_{u,\widehat\theta_t'}^{\circ}(L_u)}\le\sup_{\theta\in\Theta}\sum_{u=1}^t\log\frac{q_{u,\theta}(L_u)}{q_{u,\widehat\theta_t}(L_u)}+C=o(t)\quad\text{a.s.}
    \]
    Thus $\widehat\theta_t'$ satisfies the fitting condition for the stationary family. Moreover, the bound for the change in log density gives
    \[
        \sum_{u\ge1}\left|\log\frac{q_u(L_u)}{q_{u,\widehat\theta_{u-1}'}^{\circ}(L_u)}\right|\le C\sum_{u\ge1}(1-\varepsilon)^{u-1}\{1+\sigma_1^2(\widehat\theta_{u-1})\}<\infty\quad\text{a.s.},
    \]
    where the last series converges by the root test. The proof of Proposition~\ref{prop:auxiliary_precision}, applied to the stationary family, gives a vanishing average absolute log ratio for $q_{u,\theta_0}^{\circ}$ and $q_{u,\widehat\theta_{u-1}'}^{\circ}$. The initialization error at $\theta_0$ is also summable because $\sigma_1^2(\theta_0)<\infty$ a.s. Combining these bounds gives
    \[
        \frac1t\sum_{u=1}^{t-1}\left|\log\frac{q_{u,\theta_0}(L_u)}{q_u(L_u)}\right|\to0\quad\text{a.s.}
    \]

    Finally, the running intersection bounds the initial variance of every retained candidate. Since $f_{\nu,\gamma}\le C$,
    \[
        E_1(\theta)<1/\alpha\quad\Longrightarrow\quad\sigma_1(\theta)<\frac{C(1/\alpha-1+\lambda)}{\lambda q_1(L_1)}<\infty\quad\text{a.s.}
    \]
    Hence all $\mathcal C_t$, $t\ge1$, lie in one set with bounded initial variance. For each fixed integer $M\ge B$, coupling the initialized and stationary laws by the same innovation gives
    \[
        \sup_{\theta:\,\sigma_1^2(\theta)\le M}|\Delta_{0,t}(\theta)-\Delta_{0,t}^{\circ}(\theta)|\le\frac Ct\sum_{u=1}^t(1-\varepsilon)^{u-1}\operatorname{Lip}\{S(\cdot,R_u)-S(\cdot,R_u^*)\}\to0\quad\text{a.s.}
    \]
    The infinite series has expectation $\varepsilon^{-1}\E\operatorname{Lip}\{S(\cdot,R_1)-S(\cdot,R_1^*)\}<\infty$. The uniform likelihood and score bounds from the proof of Proposition~\ref{prop:auxiliary_precision} therefore give, on each such compact set,
    \[
        \liminf_{t\to\infty}\inf_{\theta:\,\sigma_1^2(\theta)\le M,\ |\Delta_{0,t}(\theta)-\Delta_t|\ge\delta}\frac{\log E_{t-1}(\theta)}t>0\quad\text{a.s.},\qquad\delta>0.
    \]
    These limits hold simultaneously for integer $M$ and $\delta=1,1/2,\ldots$. Choose $M$ above the square of the preceding bound on $\sigma_1(\theta)$. The separation inequality then gives, eventually a.s.,
    \[
        \mathcal C_{t-1}\ne\varnothing\quad\Longrightarrow\quad\max\{|\underline\Delta_{0,t}-\Delta_t|,|\overline\Delta_{0,t}-\Delta_t|\}\le\delta.
    \]
    Taking $\delta=1,1/2,\ldots$ and using \eqref{eq:auxiliary_cs} proves the claimed probability bound and the almost sure conclusion under eventual nonemptiness.
\end{proof}

\section{Supplementary numerical results}
\label{app:sec:num}

\subsection{Stationary comparisons at the intermediate risk levels}
\label{app:sec:ts_intermediate}

\begin{figure}[htbp]
    \centering
    \includegraphics[width=\linewidth]{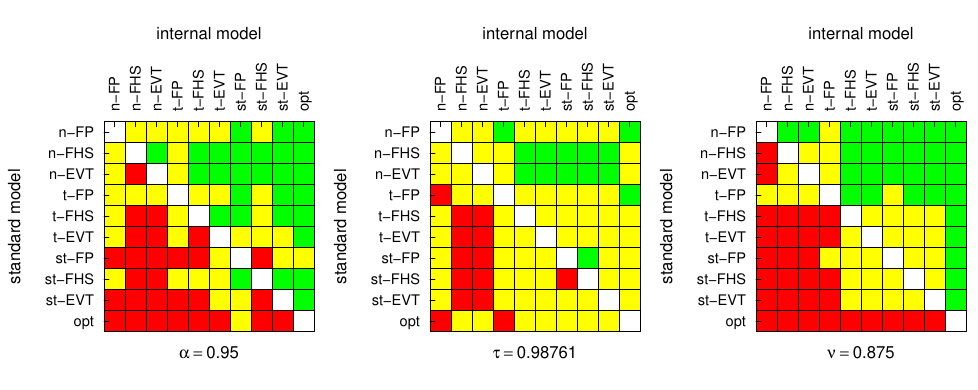}
    \caption{Overall-test heatmaps for stationary time series forecasts (left panel: $\VaR_{0.95}$, middle panel: $\ex_{0.98761}$, right panel: $(\ES_{0.875},\VaR_{0.875})$). The horizontal axis represents the internal model and the vertical axis represents the standard model.}
    \label{fig:supp-intermediate}
\end{figure}

Figure \ref{fig:supp-intermediate} gives the intermediate risk levels of Section \ref{sec:time_series}, using the same data, forecasts and likelihood confidence region as the main figures. The test resolves 28 of 45 distinct pairs for VaR, 14 for expectiles and 30 for VaR--ES. Together with the main figures, these results show that resolution changes little across levels for VaR and VaR--ES, while it decreases from 17 to 14 to nine pairs as the expectile level increases.

\subsection{Additional NASDAQ resetting comparisons}
\label{app:nasdaq_resetting}

\begin{figure}[p]
    \centering
\includegraphics[width=\textwidth,height=0.62\textheight,keepaspectratio]{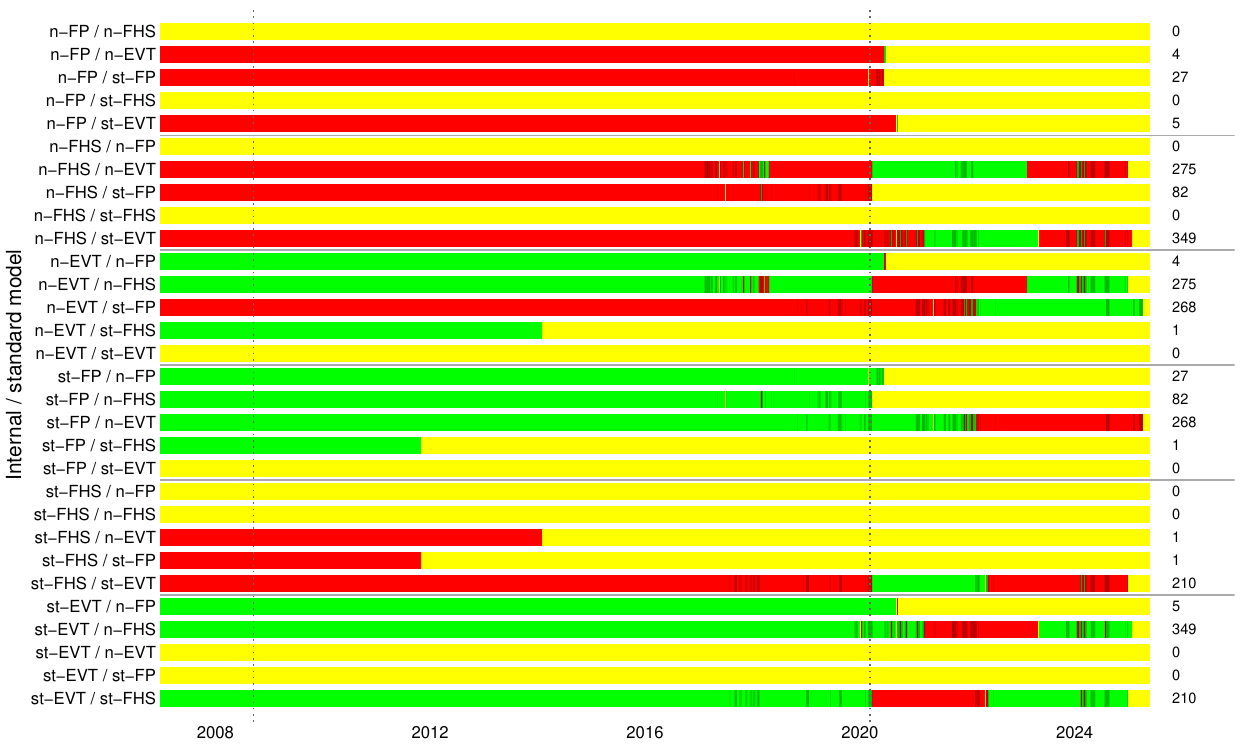}
    \caption{Resetting reports for all 30 ordered NASDAQ VaR comparisons. Rejection points are marked in gray. Each row gives the internal model followed by the standard model. Numbers give separate completed report counts. Dotted lines mark September 15, 2008 and March 11, 2020.}
    \label{fig:nasdaq-reset-VaR}
    \medskip
\includegraphics[width=\textwidth,height=0.62\textheight,keepaspectratio]{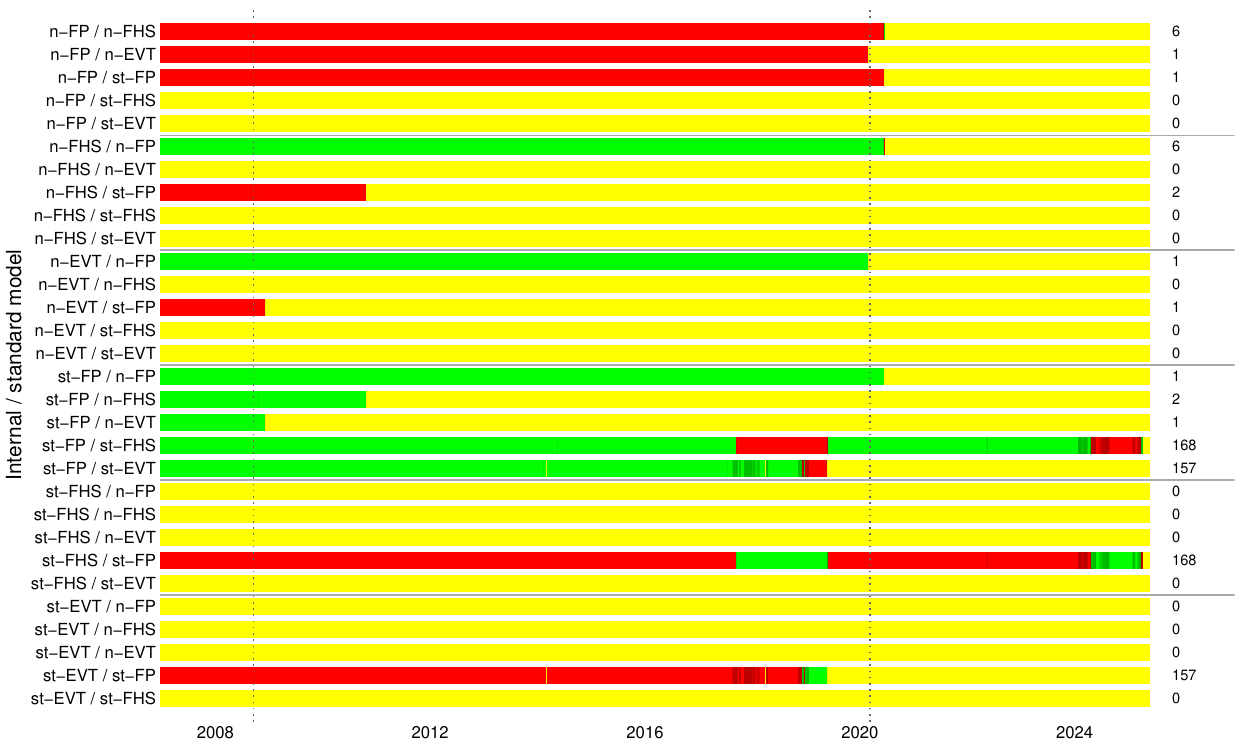}
    \caption{Resetting reports for all 30 ordered NASDAQ expectile comparisons. Rejection points are marked in gray. Each row gives the internal model followed by the standard model. Numbers give separate completed report counts. Dotted lines mark September 15, 2008 and March 11, 2020.}
    \label{fig:nasdaq-reset-EXP}
\end{figure}

Figures \ref{fig:nasdaq-reset-VaR} and \ref{fig:nasdaq-reset-EXP} display all 30 ordered comparisons for VaR and expectiles, using the same data, forecasts and likelihood confidence sequence as Figure~\ref{fig:realeprocess}. Both colors occur over different assessment intervals for seven distinct VaR pairs and three expectile pairs. For example, comparing st-FHS with st-FP for expectiles gives a green report for September 13, 2017--May 20, 2019, and a red report for May 28, 2019--May 10, 2022. The terminal comparison favors st-FP, but the resetting reports show that this overall conclusion does not extend to every assessment period.

\subsection{Replication of the Nolde--Ziegel three-zone approach}
\label{app:nz_replication}

For comparison, we apply the procedure of \cite{NZ17} to the same 5000 stationary assessment losses and forecasts. We use the Parzen HAC estimator with truncation $\lceil2\sqrt n\rceil$ and a one-sided p-value cutoff of $0.05$ for each direction, following the accompanying code. The upper panels use the scores in Example \ref{ex:2}, and the lower panels use the scores with $0$-homogeneous differences in Appendix \ref{app:sec:score0}. Exchanging the models negates the score difference and leaves its estimated standard error unchanged, so the figures have red--green symmetry.

With the original scores, the fixed-time comparisons resolve 23, 21 and 28 of 45 distinct pairs at the highest VaR, expectile and VaR--ES levels, respectively. The corresponding likelihood counts are 28, nine and 29. Thus the likelihood test resolves more VaR pairs and fewer expectile pairs on this path, with similar VaR--ES resolution. These comparisons use fixed-time asymptotic calibration at $0.05$ in each direction, whereas the likelihood test has one common anytime error allowance of $0.05$.

Changing the scoring function also changes which pairs are resolved. At the lowest expectile level, the number of resolved pairs increases from 18 to 26 under the $0$-homogeneous score, and 24 of 45 pairs change color. On this path, all color changes involve yellow; no pair is conclusively ranked in opposite directions by the two scores. The $0$-homogeneous scores do not increase resolution at every risk level. The scoring function therefore affects both the comparison criterion and the number of conclusive comparisons.

\begin{figure}[htbp]
    \centering
    \includegraphics[width=0.90\textwidth,height=0.62\textheight,keepaspectratio]{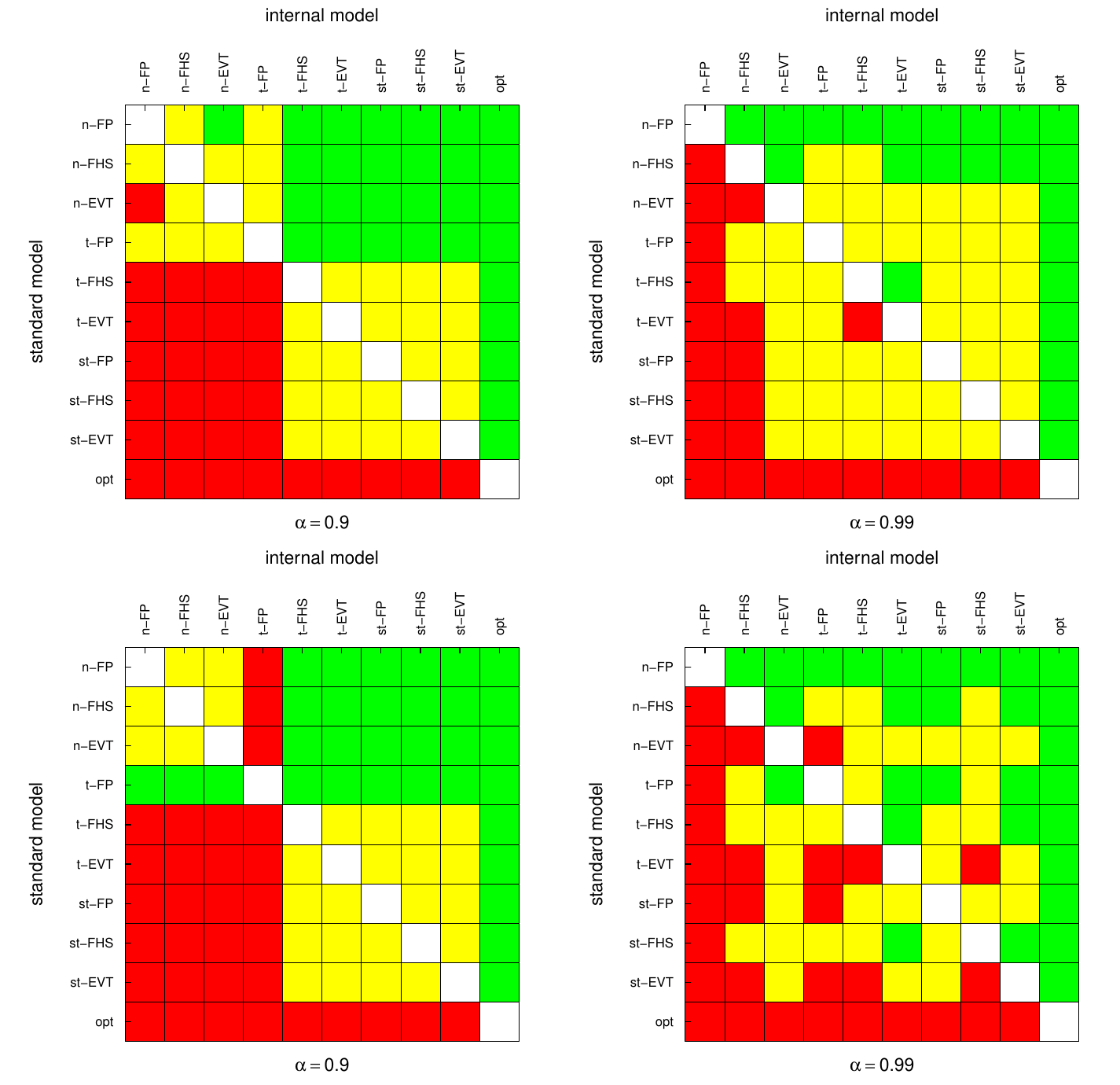}
    \caption{Replication of the Nolde--Ziegel three-zone approach for $\VaR_p$ forecasts at levels $p=0.9$ and $0.99$ for stationary time series data. The upper panels use the original score and the lower panels use the score with $0$-homogeneous differences. The horizontal axis represents the internal model and the vertical axis represents the standard model.}
    \label{fig:nz-VaR}
\end{figure}

\begin{figure}[htbp]
    \centering
    \includegraphics[width=0.90\textwidth,height=0.62\textheight,keepaspectratio]{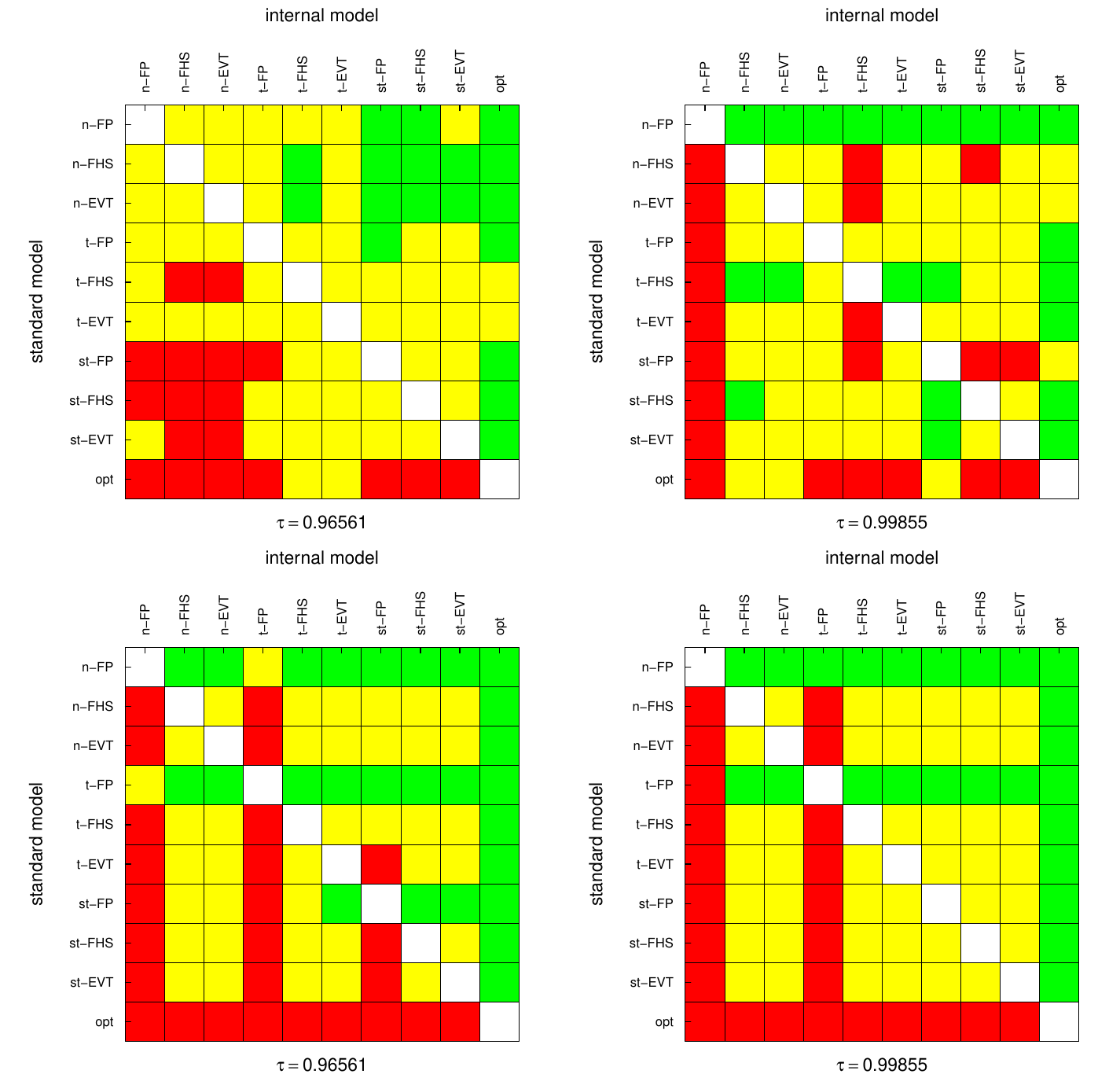}
    \caption{Replication of the Nolde--Ziegel three-zone approach for $\ex_\tau$ forecasts at levels $\tau=0.96561$ and $0.99855$ for stationary time series data. The upper panels use the original score and the lower panels use the score with $0$-homogeneous differences. The horizontal axis represents the internal model and the vertical axis represents the standard model.}
    \label{fig:nz-EXP}
\end{figure}

\begin{figure}[htbp]
    \centering
    \includegraphics[width=0.90\textwidth,height=0.62\textheight,keepaspectratio]{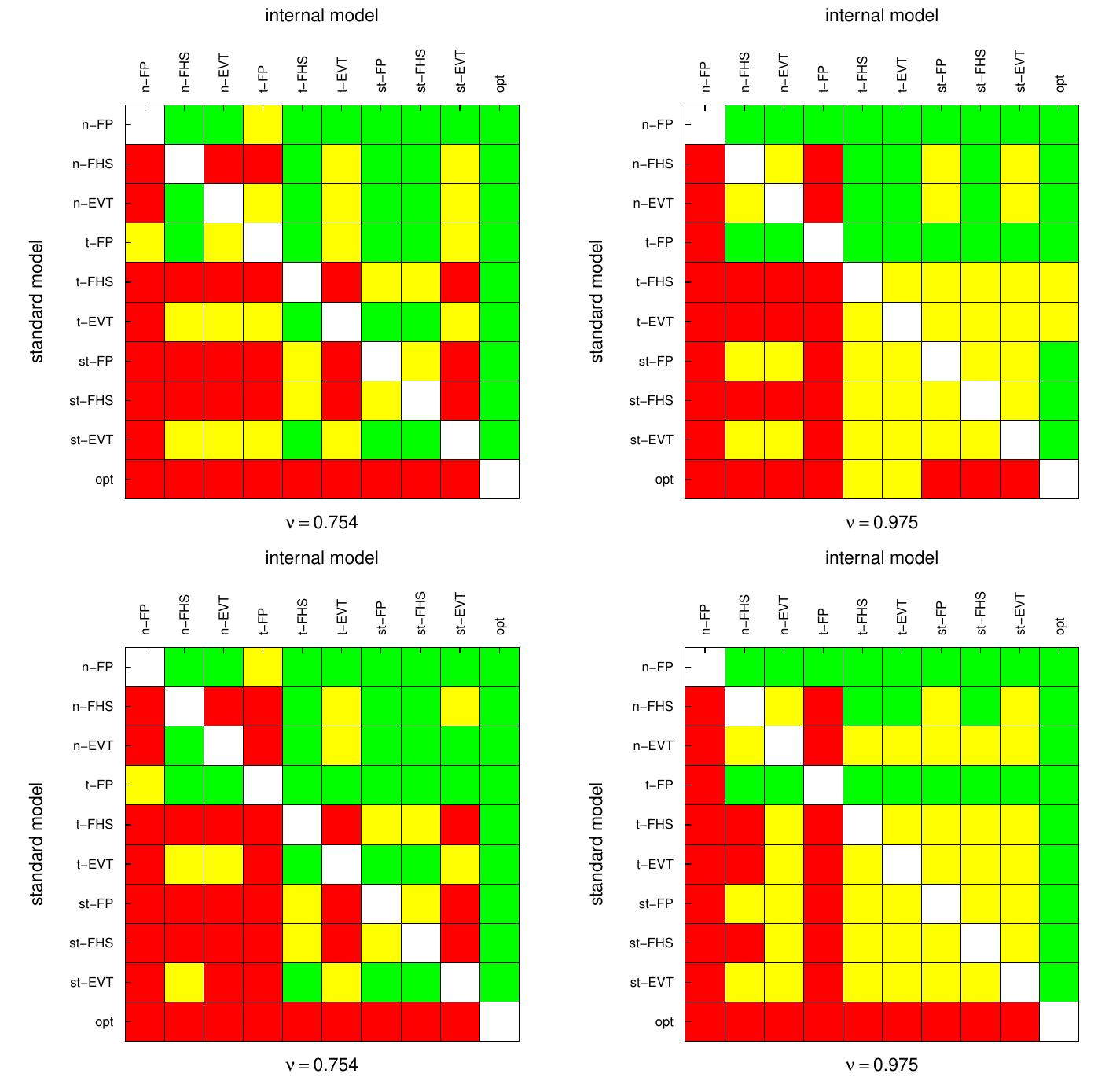}
    \caption{Replication of the Nolde--Ziegel three-zone approach for $(\ES_\nu,\VaR_\nu)$ forecasts at levels $\nu=0.754$ and $0.975$ for stationary time series data. The upper panels use the original score and the lower panels use the score with $0$-homogeneous differences. The horizontal axis represents the internal model and the vertical axis represents the standard model.}
    \label{fig:nz-VaRES}
\end{figure}

\clearpage

\end{document}